\documentclass[journal]{IEEEtran}

\usepackage{multirow}     %
\usepackage{booktabs, tabularx, tabularray}     %
\usepackage{amsmath,amssymb,amsfonts}
\usepackage{rotating}
\usepackage{adjustbox}
\usepackage{graphicx}
\usepackage{lscape}
\usepackage{algorithm}
\usepackage{algpseudocode}
\usepackage{caption, subcaption, array, cite}
\UseTblrLibrary{booktabs}

\begin{document}

\bstctlcite{IEEEexample:BSTcontrol}
\title{ESSR: An 8K@30FPS Super-Resolution Accelerator With Edge Selective Network
}

\author{\IEEEauthorblockN{Chih-Chia Hsu, and Tian-Sheuan Chang, \textit{Senior Member, IEEE}}
\thanks{This work was supported by the National Science and Technology Council, Taiwan, under Grant 111-2622-8-A49-018-SB, 110-2221-E-A49-148-MY3, and 110-2218-E-A49-015-MBK. The authors are affiliated with the Institute of Electronics, National Yang Ming Chiao Tung University, Taiwan. (e-mail: hsu880105@gmail.com, tschang@nycu.edu.tw) }%
\thanks{Manuscript received XXXX XX, 2023; revised XXXX XX, XXXX.}
}
\maketitle

\begin{abstract}
Deep learning-based super-resolution (SR) is challenging to implement in resource-constrained edge devices for resolutions beyond full HD due to its high computational complexity and memory bandwidth requirements. This paper introduces an 8K@30FPS SR accelerator with edge-selective dynamic input processing. Dynamic processing chooses the appropriate subnets for different patches based on simple input edge criteria, achieving a 50\% MAC reduction with only a 0.1dB PSNR decrease. The quality of reconstruction images is guaranteed and maximized its potential with \textit{resource adaptive model switching} even under resource constraints. In conjunction with hardware-specific refinements, the model size is reduced by 84\% to 51K, but with a decrease of less than 0.6dB PSNR. Additionally, to support dynamic processing with high utilization, this design incorporates a \textit{configurable group of layer mapping} that synergizes with the \textit{structure-friendly fusion block}, resulting in 77\% hardware utilization and up to 79\% reduction in feature SRAM access. The implementation, using the TSMC 28nm process, can achieve 8K@30FPS throughput at 800MHz with a gate count of 2749K, 0.2075W power consumption, and 4797Mpixels/J energy efficiency, exceeding previous work.

~
\end{abstract}

\section{Introduction}
\label{chapter:introduction}
Deep learning-based super-resolution (SR) has gained prominence in recent years due to its exceptional performance. The growing demand for high-resolution (HD), ultra-HD or even 8K images in various edge device applications, including surveillance, medical imaging, virtual reality and digital entertainment, underscores its importance. However, the computational demands and memory bandwidth requirements of these SR methods present significant challenges for resource-constrained edge devices aiming for real-time SR execution. Consequently, there is a pressing need for efficient hardware accelerators.

Various hardware accelerators have been proposed in recent years~\cite{ECNN, HPAN, SRNPU, SRNPU2022, ACNPU} for HD applications. However, due to the extensive computational demands and significant memory bandwidth requirements, many existing super-resolution accelerators opt for simplistic and extremely lightweight models, such as FSRCNN \cite{FSRCNN} or 1-D convolution~\cite{HPAN}, as their backbone. This often results in a compromise in both performance and perceptual quality. The only exception is eCNN~\cite{ECNN}, which employs a heavy, yet simple structure that consumes a substantial area but fails to deliver superior quality. Besides, these designs only consider PSNR-oriented methods that maximize the output PSNR to be close to the ground truth. This weights all pixel differences equally and could become blurry on the highly textured part. None have considered perceptual-oriented methods like GAN~\cite{SRGAN, ESRGAN, RealESRGAN, GAN1} that prioritize the enhancement of reconstructed images based on human perception. This will lead to sharper and more textured images, but could introduce artifacts and noise not present in the original image.

To enable real-time execution, various lightweight methods have been introduced.  These include techniques such as asymmetric convolution \cite{ACNet, ACNPU}, depthwise convolution~\cite{mobilenets}, parameter sharing strategies \cite{ParameterSharing1}, and knowledge distillation mechanisms \cite{RFDN}. Nevertheless, only a handful of lightweight super-resolution networks take hardware design considerations into account \cite{HPAN, SRNPU, ACNPU}.

Moving beyond traditional methods, dynamic processing techniques \cite{classsr, ARM, dynamic2022, SRNPU, wang2023classification} have emerged as a promising strategy to strike a balance between complexity and performance. Such methods deploy lightweight networks for simpler inputs and more resource-intensive networks for challenging inputs~\cite{classsr}. For example, SRNPU~\cite{SRNPU} incorporates a classification network to select a subnet; however, its decision-making is not adaptive to the constraints of available hardware resources. Furthermore, its hardware utilization is only 31.2\% for the smaller network. CDNSR~\cite{wang2023classification} also needs a classification network to select three different sizes of SR networks. In contrast, ARM~\cite{ARM} uses edge-to-PSNR lookup tables to choose one of its four subnets. Unlike SRNPU and ClassSR \cite{classsr}, which are burdened with a high parameter count due to their multiple independent subnets, ARM utilizes shared weights between subnets. However, the edge-to-PSNR table in ARM demands extra area and proves challenging to adapt within the confines of available hardware resources. \cite{dynamic2022} allocates bits adaptively based on the local contents of an input image, which is not friendly for hardware design.  In summary, existing works need complex subnet decisions, have no quality guarantee under limited resources, and often lead to low hardware utilization.

Briefly, existing SR designs face three challenges: simple yet effective dynamic processing, improved perceptual quality with extremely lightweight models (approximately 50K in model size), and efficient hardware that supports 8K outputs.
Motivated by these challenges, this paper proposes an \textit{edge selective SR network} (ESSR) and its design. To address the first challenge, we introduce an input edge threshold that adaptively selects a proper subnet. This approach saves 50\% of multiply-accumulate (MAC) operations with only a 0.1dB drop. All subnets share the same weights, resulting in a smaller model size and a consistent hardware structure. For the second challenge, we suggest hardware-oriented modifications to a heavy but high PSNR model, preserving its quality while reducing its parameters by 84\% to 51K and its MACs by 83\%. 
We further enhance the model with perceptual-oriented training to emphasize texture and details. 
To tackle the third challenge, we introduce \textit{resource adaptive model switching}, ensuring minimum image quality while maximizing potential quality under resource constraints. Furthermore, we employ a \textit{configurable group of layer mapping}, which works in tandem with the proposed \textit{Structure-Friendly Fusion Block}, achieving 77\% hardware utilization and reducing feature SRAM accesses by up to 79\%. The final hardware implementation can achieve 8K@30FPS with an energy efficiency of 4797Mpixels/J for x4 scaling.

The remainder of the paper is organized as follows. Section II first introduces the proposed edge-selective dynamic input processing. Section III reviews the adopted network and its hardware-oriented modifications. The hardware design is presented in Section IV. The experimental results are provided in Section V. Finally, the paper is concluded in Section VI.

\section{Edge Selective Dynamic Input Processing}
\label{sec: Edge Selective Dynamic Input Processing}
\subsection{Inference of edge selective dynamic input processing}

\begin{figure}[tb]
    \centering
    \includegraphics[height=!,width=1\linewidth,keepaspectratio=true]{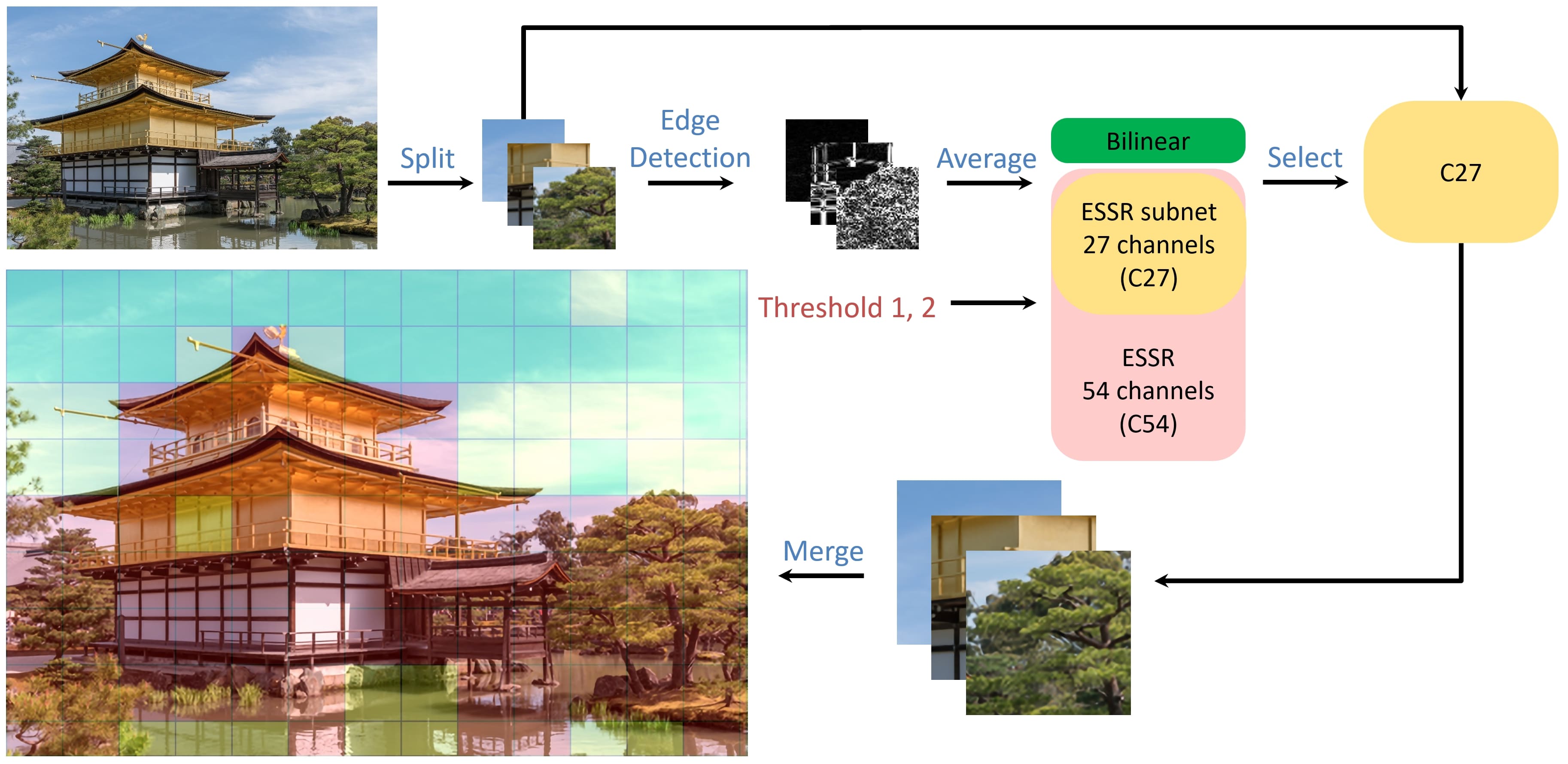}
    \caption {Inference of the edge selective dynamic input processing. \textbf{Green patch}: Bilinear interpolation. \textbf{Yellow patch}: C27. \textbf{Red patch}: C54.}
    \label{dynamic inference}
\end{figure}

As depicted in Fig.~\ref{dynamic inference}, we detail the proposed dynamic inference procedure. Initially, the LR image is segmented into overlapping patches. Subsequently, an edge score is computed for each patch, guiding the selection of the appropriate subnet based on this score. We have a trio of subnets to choose from: bilinear, C27, and C54 (Cxx is the channel number of the model). Notably, C54 leverages the full channel capacity of ESSR (shown in the next Section), whereas C27 employs just half. For the illustrated inference scenario, the choice falls on C27, into which the LR patch is inputted to generate the super-resolution counterpart. In the final stage, these super-resolution patches are seamlessly fused to render the complete super-resolution image.

To compute the edge score, we begin by extracting the luminance component from an LR patch. Edges are then detected using a $3\times3$ Laplacian filter, with the output's absolute value clamped within the range of 0 to 255. The average value of this output serves as the edge score.

\subsection{Proposed subnet decision: subnet types}
\label{section: subnet types}
Edge information serves as a valuable metric for selecting a network of varying complexity. ARM~\cite{ARM} employs an edge-to-PSNR table to predict the PSNR for each subnet and then chooses the appropriate subnet based on a balance between predicted PSNR and MAC operations. However, this strategy demands extra storage for the edge-to-PSNR table and introduces additional computational overhead for the tradeoff function. In particular, while ARM integrates four subnets, Bilinear, C16, C36, and C56, our experiments reveal that only three are actively employed in real-world scenarios.

\begin{figure}[tb]
    \centering
    \includegraphics[height=!,width=0.9\linewidth,keepaspectratio=true]{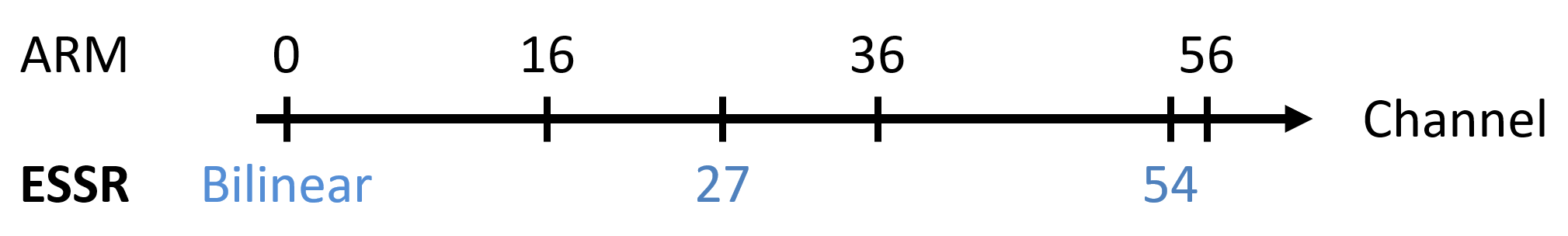}
    \caption {The subnet types of the ESSR when compared to ARM.}
    \label{dynamic_subnet_types}
\end{figure}

\begin{figure}[tb]
\centering
\begin{subfigure}[b]{0.45\linewidth}
    \centering
    \includegraphics[height=!,width=0.8\linewidth,keepaspectratio=true]{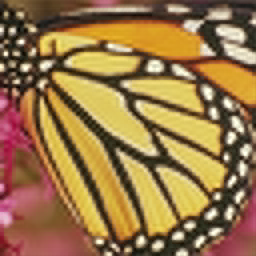}
    \caption {C16 output.}
    \label{C16_butterfly}
    \end{subfigure}
    \hfill
\begin{subfigure}[b]{0.45\linewidth}
    \centering
    \includegraphics[height=!,width=0.8\linewidth,keepaspectratio=true]{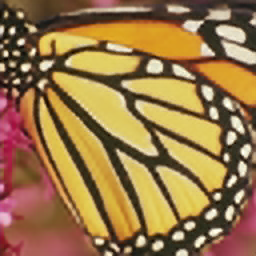}
    \caption {C27 output.}
    \label{C27_butterfly}
    \end{subfigure}
    \hfill
    \caption{The comparison of the C16 and C27. The image is from Set5 butterfly.}
    \label{compare C16 C27}
\end{figure}

Drawing from these observations, we advocate for the use of three subnets over four, as depicted in Fig.~\ref{dynamic_subnet_types}. This streamlined approach simplifies decision making and improves performance. In terms of subnet depth: 
\begin{itemize}
    \item The simplest, bilinear interpolation, is our preferred method for processing patches with plain backgrounds.
    \item For the most complex subnet, we favor C54 over C56. This choice is rooted in the fact that 54 is a multiple of 9, corresponding to the count of 3×3 depth-wise convolution processing elements (PEs). This alignment eases hardware design, particularly in critical path timing and modularization.
    \item As for the intermediate subnet, we gravitate towards C27, positioned between C16 and C36 in ARM. This preference arises due to the sub-par performance of C16 within our ESSR model, evident from pronounced 4×4 blocks in the C16 output. In stark contrast, C27 is devoid of this issue, as illustrated in Fig.~\ref{compare C16 C27}. Moreover, the value 27, being half of 54 and also divisible by nine, is optimal for hardware deployment.
\end{itemize}

\begin{figure*}[tb]
\centering
\begin{subfigure}[b]{0.3\textwidth}
    \centering
    \includegraphics[height=!,width=1.0\linewidth,keepaspectratio=true]{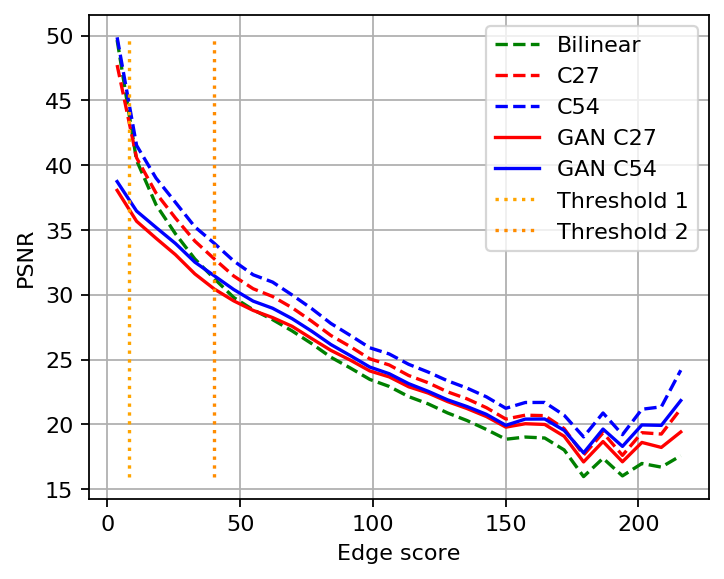}
    \caption {PSNR vs edge score.}
    \label{PSNR vs edge}
    \end{subfigure}
    \hfill
\begin{subfigure}[b]{0.3\textwidth}
    \centering
    \includegraphics[height=!,width=1.0\linewidth,keepaspectratio=true]{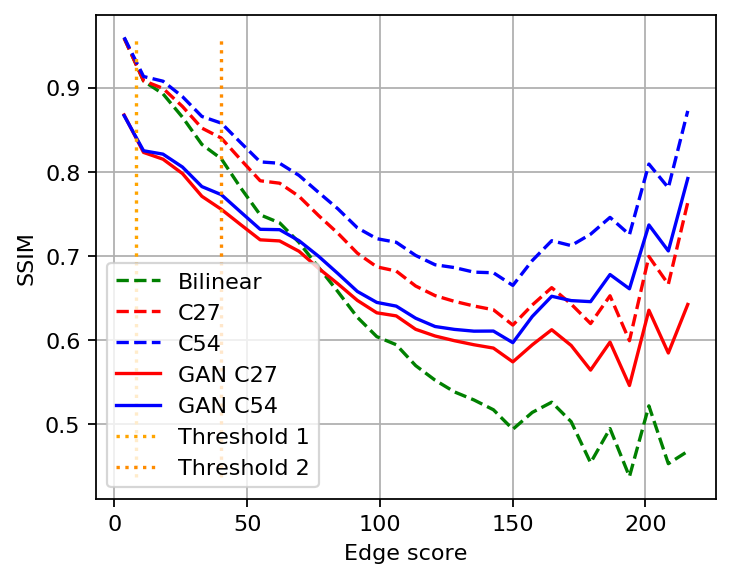}
    \caption {SSIM vs edge score.}
    \label{SSIM vs edge}
    \end{subfigure}
    \hfill
\begin{subfigure}[b]{0.3\textwidth}
    \centering
    \includegraphics[height=!,width=1.0\linewidth,keepaspectratio=true]{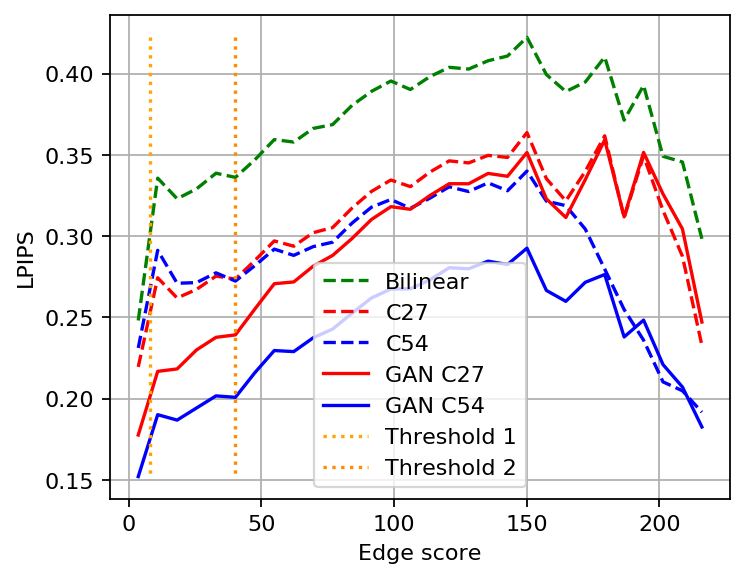}
    \caption {LPIPS vs edge score.}
    \label{LPIPS vs edge}
    \end{subfigure}
    \hfill
    \caption{The relation between edge score and bilinear, C27, C54, GAN-based C27, and GAN-based C54. Higher values of PSNR and SSIM indicate better performance, while lower values of LPIPS indicate better performance in terms of perceptual similarity.}
    \label{PSNR SSIM LPIPS vs Edge}
\end{figure*}

\subsection{Proposed subnet decision: input edge threshold}
\label{section: input edge threshold}
\begin{figure}[tb]
    \centering
    \includegraphics[height=!,width=0.9\linewidth,keepaspectratio=true]{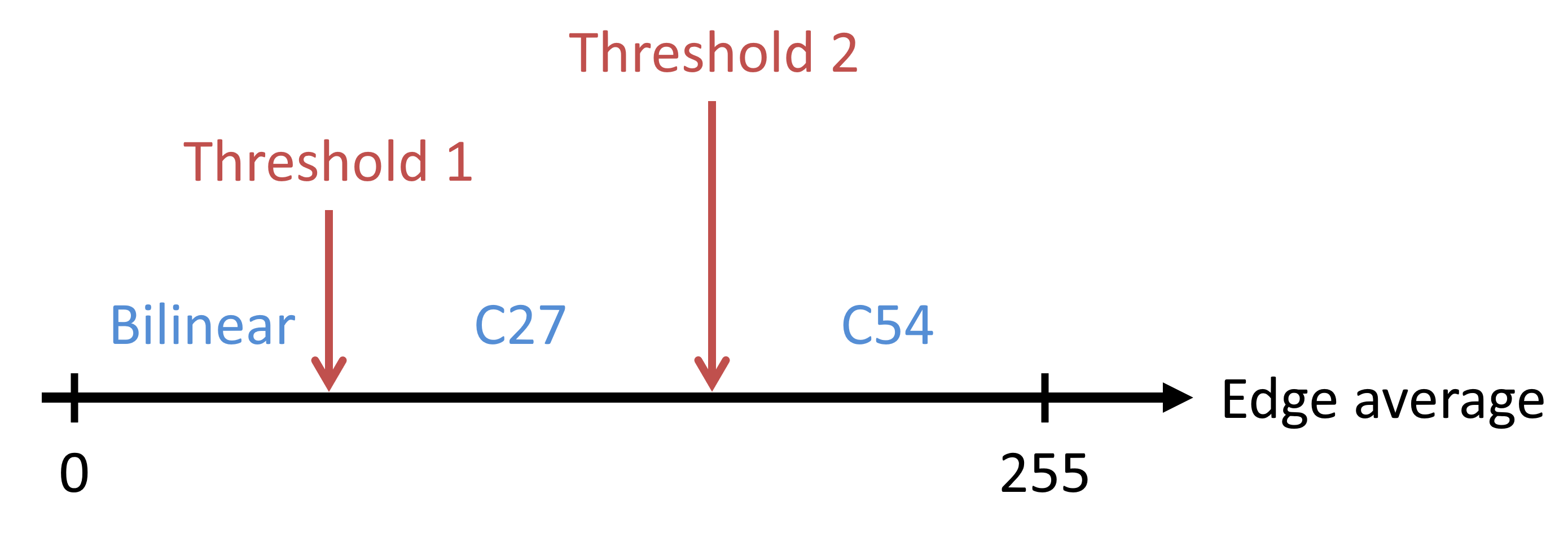}
    \caption {Proposed subnet decision with the input edge threshold .}
    \label{input edge threshold subnet decision}
\end{figure}

While ARM employs the edge-to-PSNR metric to determine the appropriate subnet, it's worth noting that a higher PSNR doesn't always correlate with better perceptual quality. Consequently, we also evaluate the edge-to-SSIM and edge-to-LPIPS metrics. An analysis of the results, as shown in Fig.~\ref{PSNR SSIM LPIPS vs Edge}, reveals that the five models exhibit comparable performance with respect to PSNR and SSIM. However, there are pronounced disparities in LPIPS scores, particularly between perceptual-based models and bilinear interpolation. Given these observations, coupled with the objective of conserving MAC operations, we introduce an optimized strategy termed the \textit{subnet decision with the input edge threshold}, illustrated in Fig.~\ref{input edge threshold subnet decision}.

In our proposed method, we define two thresholds: \( threshold_{1} \), set between bilinear interpolation and C27, and \( threshold_{2} \), situated between C27 and C54. By assigning values \( threshold_{1} = 8 \) and \( threshold_{2} = 40 \), we manage to achieve an approximate savings of 50\% in MAC operations in the Test8K dataset, with a marginal 0.1dB PSNR decrease relative to the non-edge selective approach. The introduction of these input edge thresholds obviates the need for three edge-to-PSNR lookup tables, thereby simplifying the decision-making mechanism and eliminating the computation associated with the trade-off function. This is also easily adjusted according to the available resources, as indicated below.

\section{Hardware Friendly Approaches Based on RLFN}
\subsection{Reference model}
\begin{figure}[tb]
\centering
\includegraphics[width=0.7\linewidth,keepaspectratio=true]{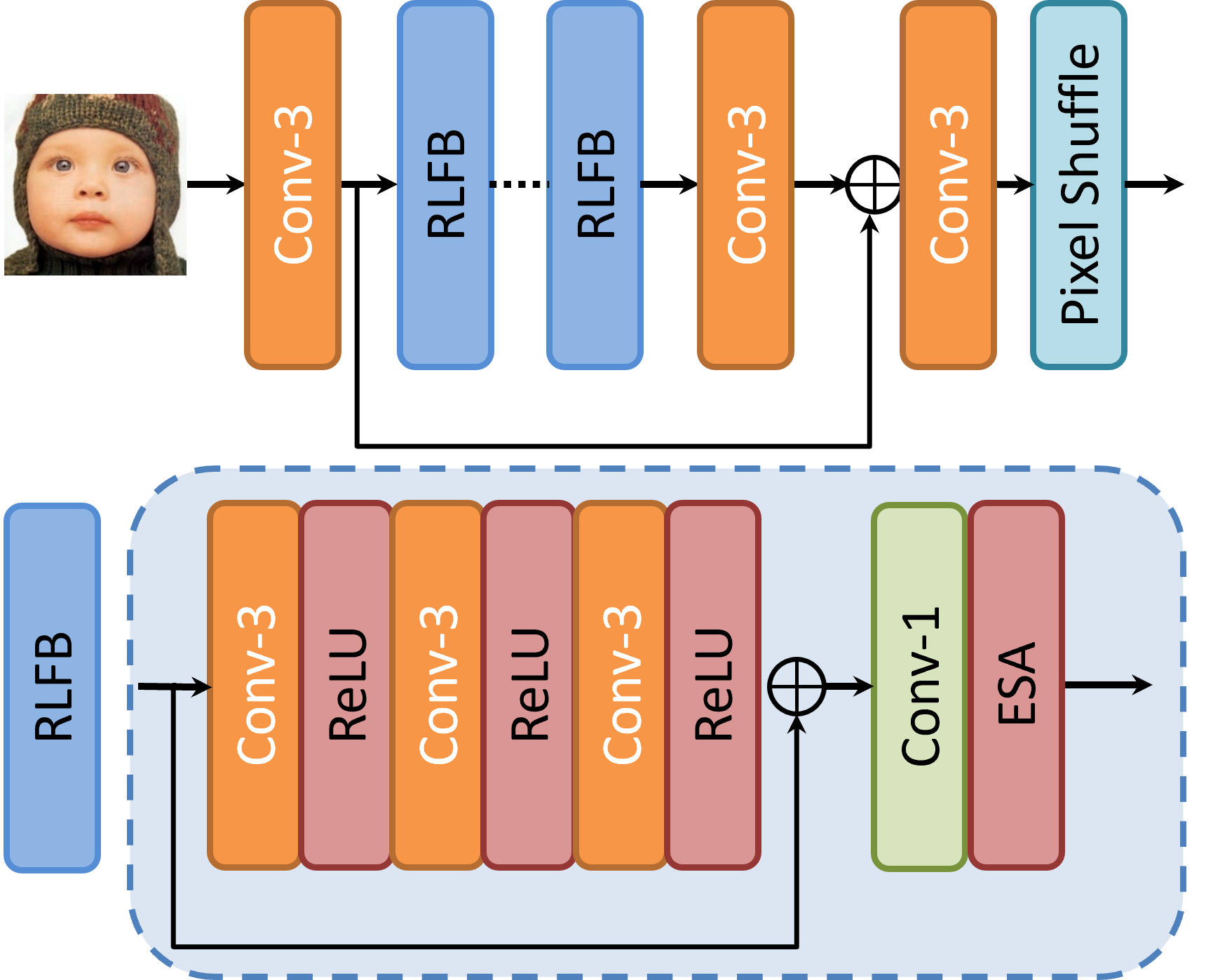}
    \caption{The network architecture of the RLFN. The conv-\textit{k} is \textit{kxk} convolution. }
    \label{RLFN network architecture.}
\end{figure}

We have chosen RLFN \cite{RLFN} as our reference model, given its commendable achievement of securing first place in the NTIRE 2022 efficient super-resolution challenge \cite{NTIRE2022}. The network architecture of RLFN is illustrated in Fig.~\ref{RLFN network architecture.}. While RLFN has a structured and intuitive network design, it presents certain complications for hardware implementation. These challenges include a substantial count of MAC operations, the inclusion of attention blocks, and an excessive amount of shortcuts.

\subsection{Hardware friendly modifications}
\begin{figure}[tb]
    \centering
    \includegraphics[height=!,width=1\linewidth,keepaspectratio=true]{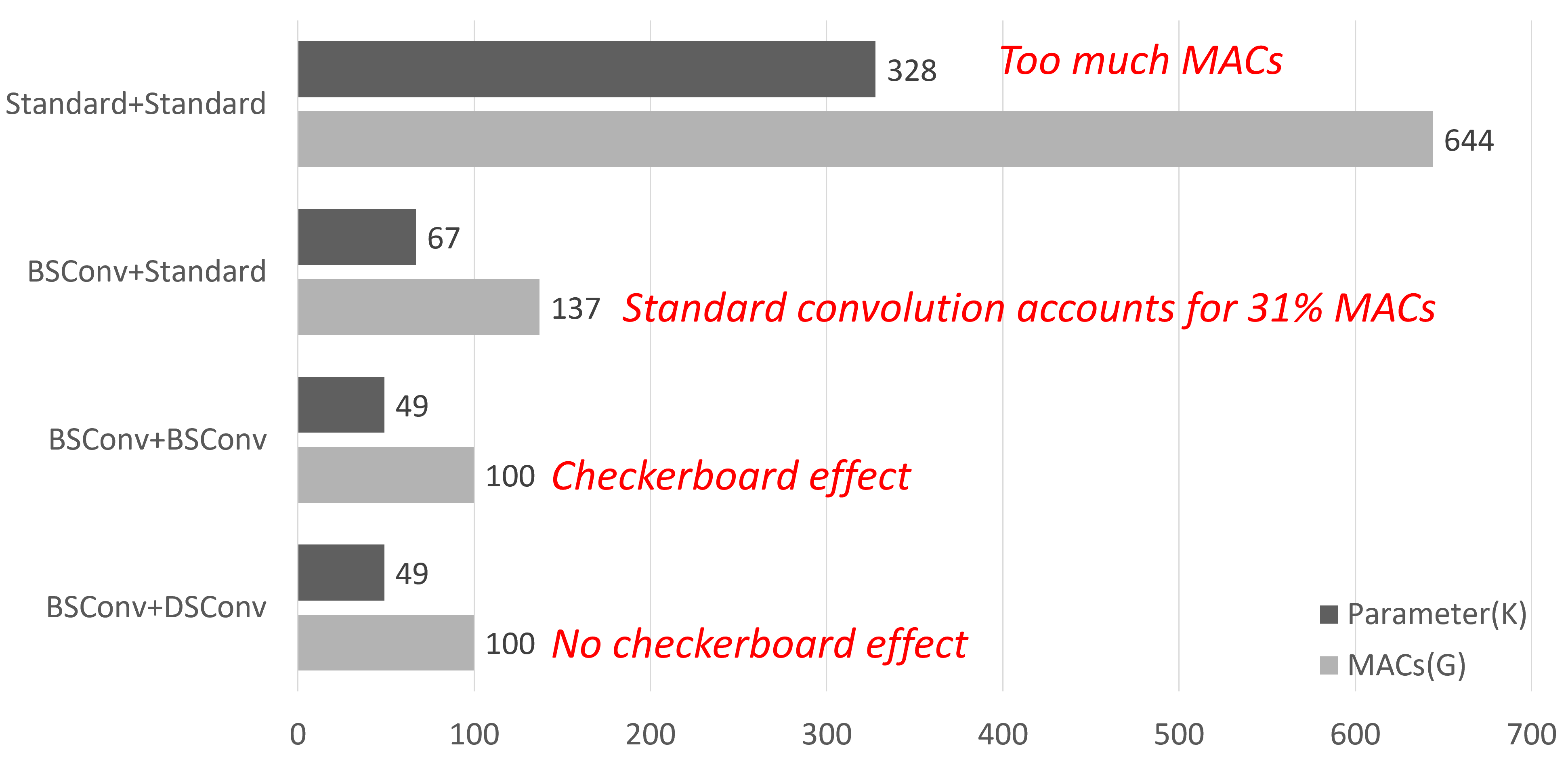}
    \caption {The comparison of different convolution methods.}
    \label{compare convolution parameters and MACs}
\end{figure}

In our pursuit of a more hardware-friendly design, we applied several modifications to the original model:

\begin{enumerate}
    \item \textbf{Global Shortcut Removal:} The global shortcut requires storing feature maps for later use. By eliminating this, we save 33\% of the feature SRAM sizes. This removal will result in performance loss, 0.36dB drop. This drop will be reduced with the following proposed approaches.

    \item \textbf{Elimination of Enhanced Spatial Attention (ESA):} While RLFN incorporates ESA (denoted as ESA in the figure) to improve its modeling capabilities, its impact is limited—a mere 0.04dB increase in PSNR for the B100 dataset. Furthermore, ESA's integration of pooling layers, dual shortcuts, sigmoid function, and element-wise multiplication is not good for efficient hardware implementation. Consequently, we decided to remove this block.

    \item \textbf{Convolution Modifications:} To reduce the number of MAC operations, we substituted the standard 3×3 convolutions in RLFN with depthwise convolution variations, specifically BSConv or DSConv, as depicted in Fig.~\ref{ESSR}. Fig.~\ref{compare convolution parameters and MACs} contrasts various convolution techniques in relation to their parameters and MAC counts. An initial strategy involved replacing conv-3 with BSConv, excluding the final upsampling convolution, to avoid substantial performance degradation. However, the upsampling convolution still represented a large amount of 31\% of total MACs. The replacement of upsampling with BSConv still led to a significant decline in PSNR and induced a checkerboard pattern in the output visuals. This anomaly is attributed to BSConv's structure: a preceding 1×1 convolution followed by a 3×3 depth-wise convolution. This latter convolution layer inadvertently introduced anomalous pixels, after pixel-shuffling, manifested as the checkerboard pattern. Our solution utilizes DSConv for upsampling, which sequences a 3×3 depth-wise convolution ahead of a 1×1 convolution, resulting in commendable PSNR scores, a compact model, and reduced MAC operations.

    \item \textbf{structure-friendly fusion blocks (SFBs):} This is tailored to align with our hardware specifications, which will be elaborated upon in the next Section.
\end{enumerate}

\subsection{Proposed Network: ESSR}
\label{Proposed Network: ESSR}
\begin{figure}[tb]
    \centering
    \includegraphics[height=!,width=0.8\linewidth,keepaspectratio=true]{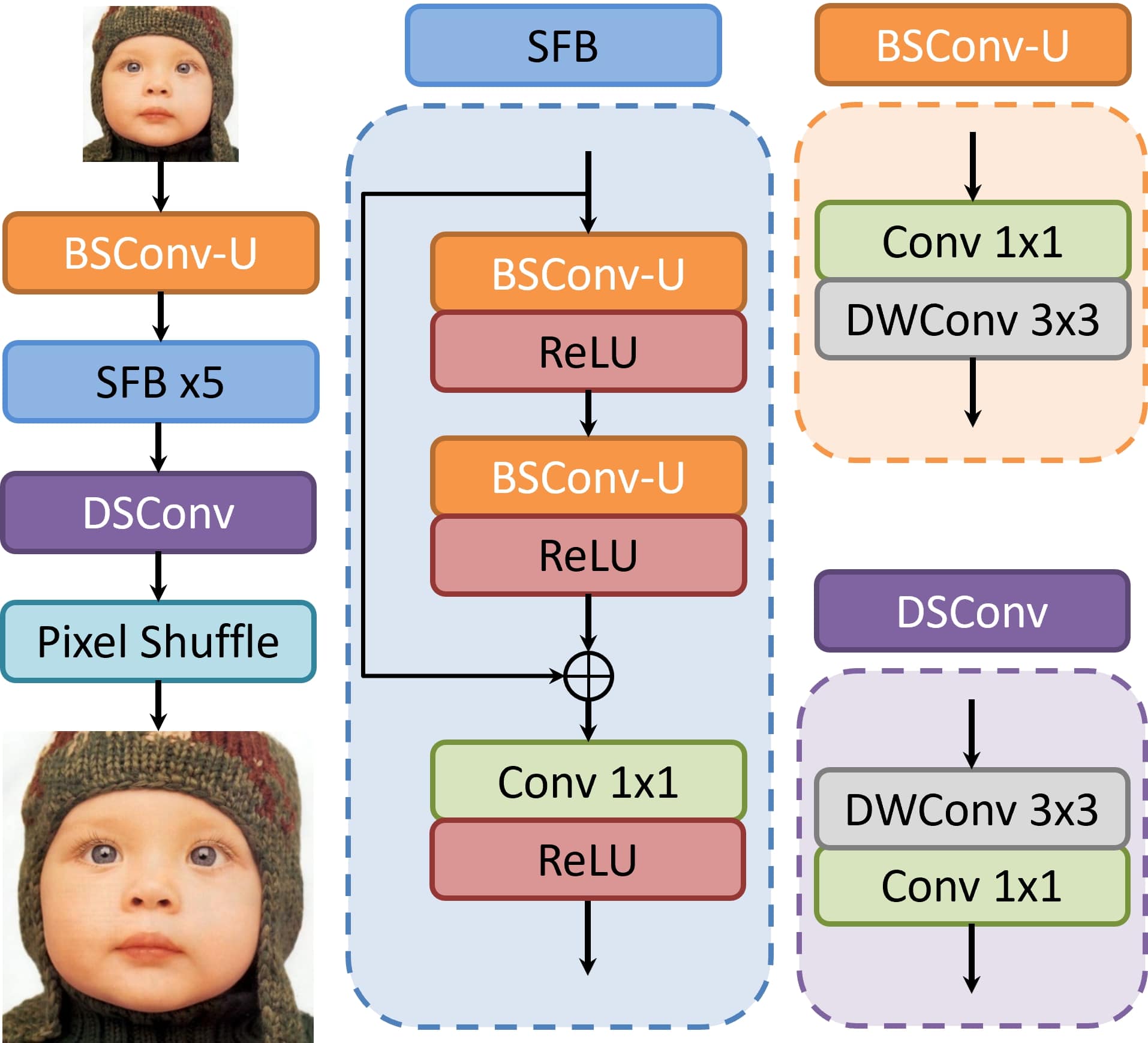}
    \caption {Proposed Edge Selective Super-Resolution network (ESSR) architecture. Conv 1×1 denotes 1×1 point-wise convolution and DWConv 3×3 denotes 3×3 depth-wise convolution.}
    \label{ESSR}
\end{figure}

The overall architecture of the ESSR is shown in Fig.~\ref{ESSR}. The ESSR architecture consists of three main parts: the first feature extraction convolution, SFBs, and the reconstruction module. This network can achieve good performance and is easy to implement in hardware. Compared to a pruned RLFN (RLFB numbers from 6 to 4, channel number from 52 to 46) for fair comparison, ESSR reduces 84\% of parameters and 83\% of MACs with a PSNR drop of less than 0.6dB.

\section{Proposed Hardware}
\label{chapter:hardware}
While the ESSR network is tailored to ease hardware design, the introduction of dynamic processing presents certain challenges. Specifically, it can lead to inconsistent image quality due to constrained hardware resources and diminished hardware utilization as a result of switching between various subnets. 

To avoid the quality issue, we introduce \textit{resource adaptive model switching}. This approach dynamically adjusts the threshold in line with the available computational resources in additional to the input edge. The goal is to ensure both baseline image quality and the highest feasible quality, all while adhering to real-time processing requirements.

To increase hardware utilization, we present the \textit{group of layer neural processing unit (GLNPU)}. Unlike traditional layer-by-layer mapping, GLNPU concurrently maps groups of layers to the hardware. This design allows parallel layer execution, enhancing hardware parallelism. Additionally, it mitigates memory access demands between layers. The efficacy of this approach is further amplified when combined with SFBs, especially for C27.

\subsection{Resource adaptive model switching}
\label{section: Resource adaptive model switching}
To achieve the specification of 8K@30FPS and accommodate fluctuating image attributes, Algorithm~\ref{alg: Resource adaptive model switching} explains the \textit{resource adaptive model switching} mechanism. This approach ensures fluid transitions between the bilinear, C27, and C54 models, all while adhering to defined temporal and computational complexity boundaries. The parameters in the algorithm are decided based on Test8K.

In the proposed algorithm:
\begin{itemize}
    \item The initial condition, wherein the number of C54 patches per second exceeds 25,500, establishes a computational ceiling, guaranteeing a steady 30FPS frame rate. Even with this constraint in place, the algorithm ensures a baseline quality by resorting to the C27 model for any residual patches. This dual constraint system facilitates the attainment of the desired throughput without compromising image quality.
    
    \item The subsequent condition supervises the allocation of C54 patches within individual frames and dynamically modulates the values of \( threshold_{1} \) and \( threshold_{2} \). If a frame utilizes a significant number of C54 patches, we increase the probability of assigning subsequent patches to the bilinear or C27 models by adjusting the values of $threshold_{1}$ and $threshold_{2}$ for complexity and real-time constraints. On the contrary, if the allocation of C54 patches is sparse for a frame, the thresholds are lowered to favor superior quality. This adaptive thresholding enables optimal hardware resource utilization, ensuring consistent throughput, and culminating in enhanced overall system performance.
\end{itemize}

\begin{algorithm}[tb]
\caption{Resource adaptive model switching}
\label{alg: Resource adaptive model switching}
\begin{algorithmic}
\State \textbf{Target:} 8K@30FPS
\State $threshold_{1} \gets 8$
\State $threshold_{2} \gets 40$
\While{not finish}
    \If{\#C54 per second $> 25500$}
        \State Rest of the patches run with C27
    \Else
        \If{\#C54 per frame $> 1000$}
            \State $threshold_{1} \gets threshold_{1} + 1$
            \State $threshold_{2} \gets threshold_{2} + 5$
        \ElsIf{\#C54 per frame $< 700$}
            \State $threshold_{1} \gets threshold_{1} - 1$
            \State $threshold_{2} \gets threshold_{2} - 5$
        \EndIf
    \EndIf
\EndWhile
\end{algorithmic}
\end{algorithm}

\begin{figure*}[!t]
    \centering
    
    \begin{minipage}{0.48\linewidth}
        \centering
        \includegraphics[width=\linewidth,keepaspectratio=true]{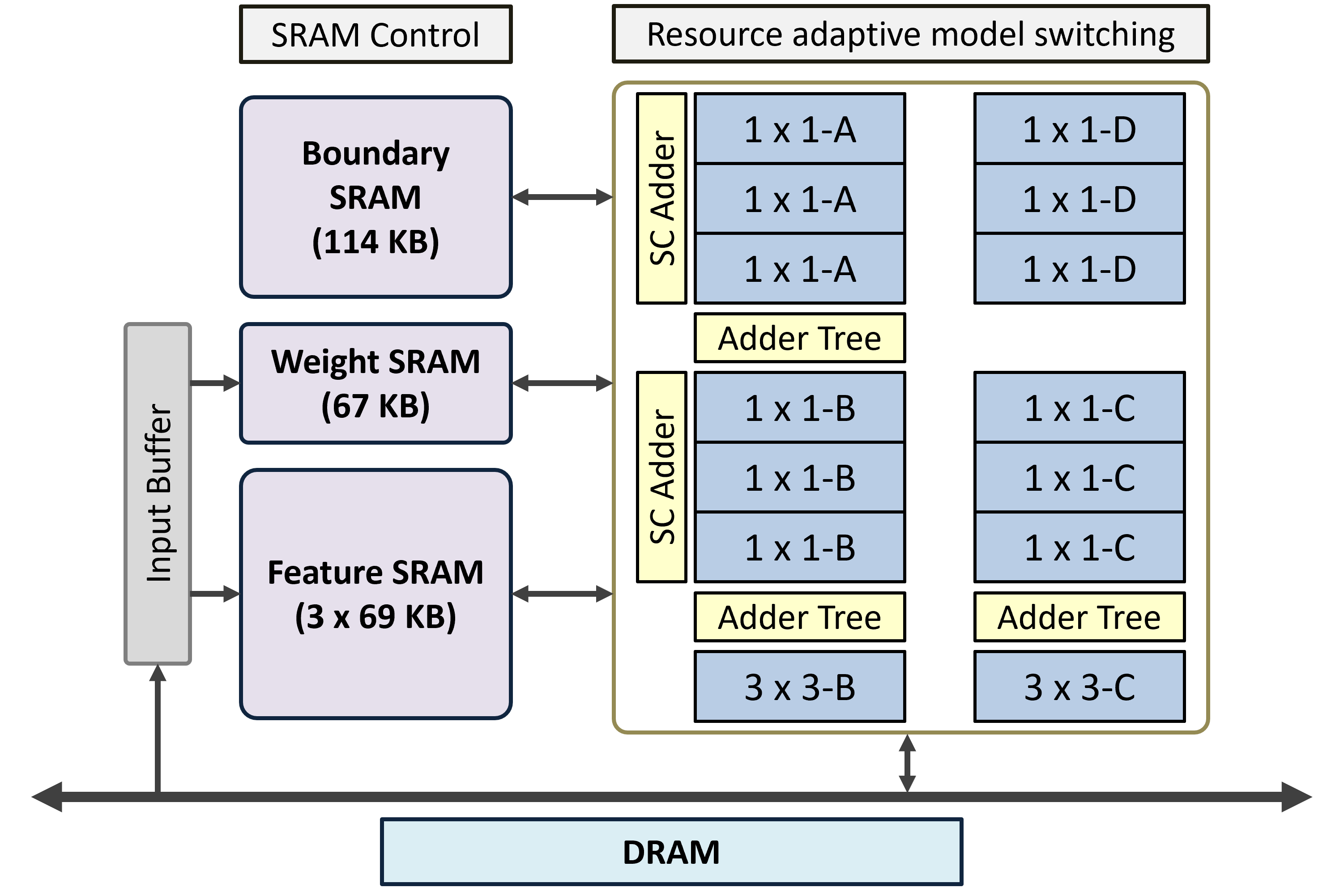}
        \caption{The proposed system architecture.}
        \label{HW_overall_arch_v3}
    \end{minipage}
    \hfill
    \begin{minipage}{0.48\linewidth}
        \centering
        \includegraphics[width=\linewidth,keepaspectratio=true]{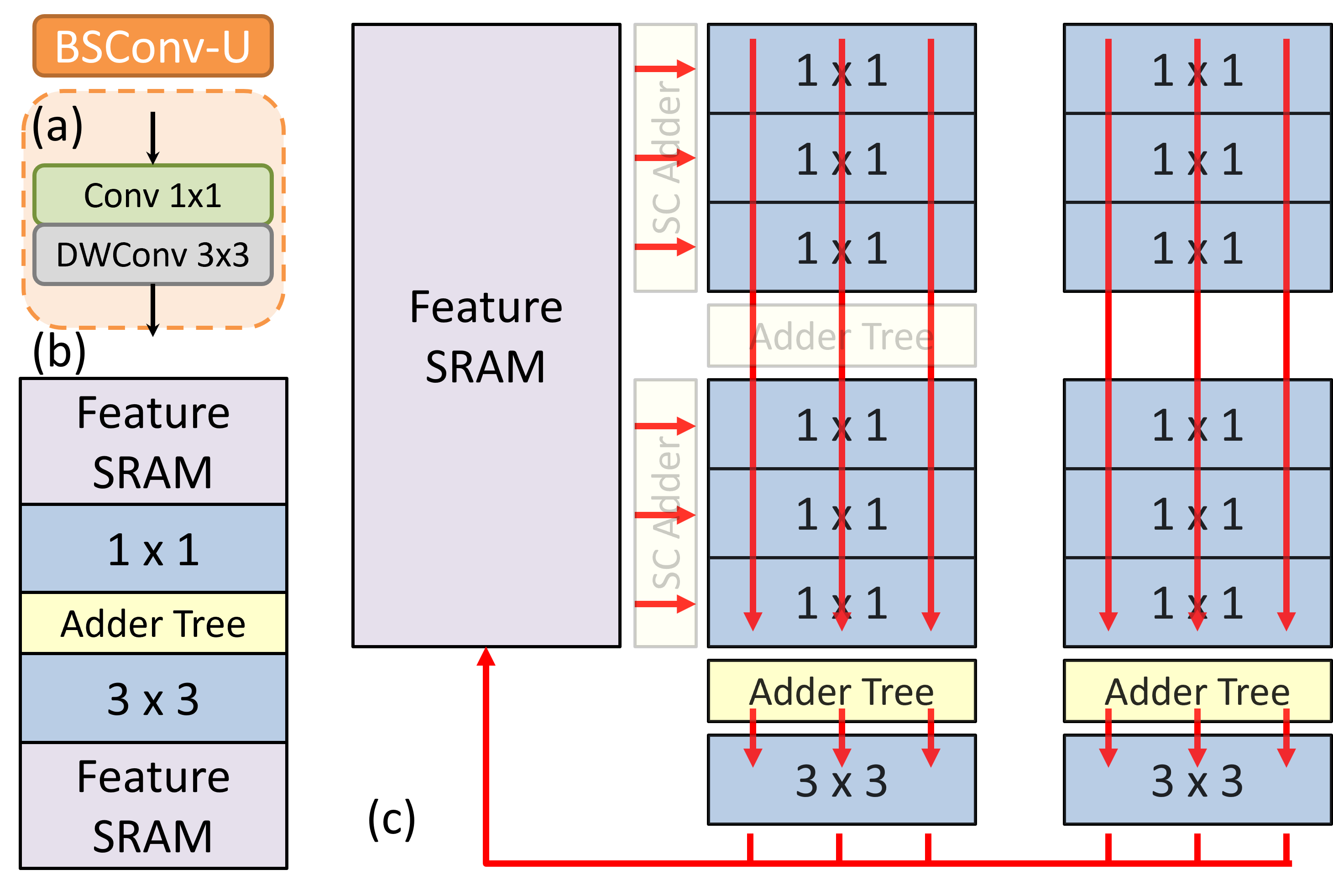}
        \caption{The dataflow of the BSConv in the C54 model. Best viewed in colors.}
        \label{HW_CLM_BSConv}
    \end{minipage}
    
    \vspace{1em}  %
    
    \begin{minipage}{0.48\linewidth}
        \centering
        \includegraphics[width=\linewidth,keepaspectratio=true]{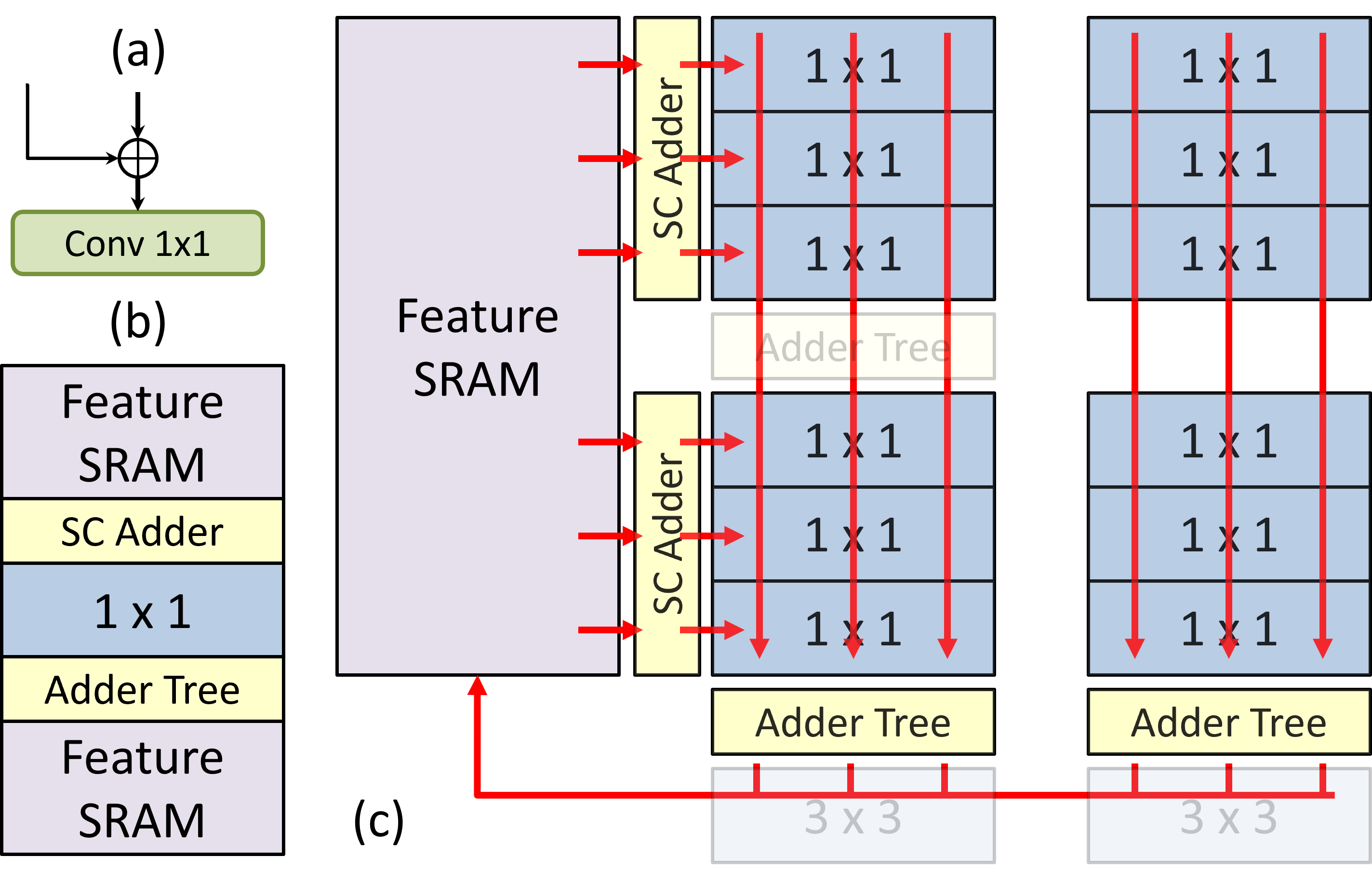}
        \caption{The dataflow of the adding shortcut with 1×1 convolution in the C54 model. Best viewed in colors.}
        \label{HW_CLM_SC1x1}
    \end{minipage}
    \hfill
    \begin{minipage}{0.48\linewidth}
        \centering
        \includegraphics[width=\linewidth,keepaspectratio=true]{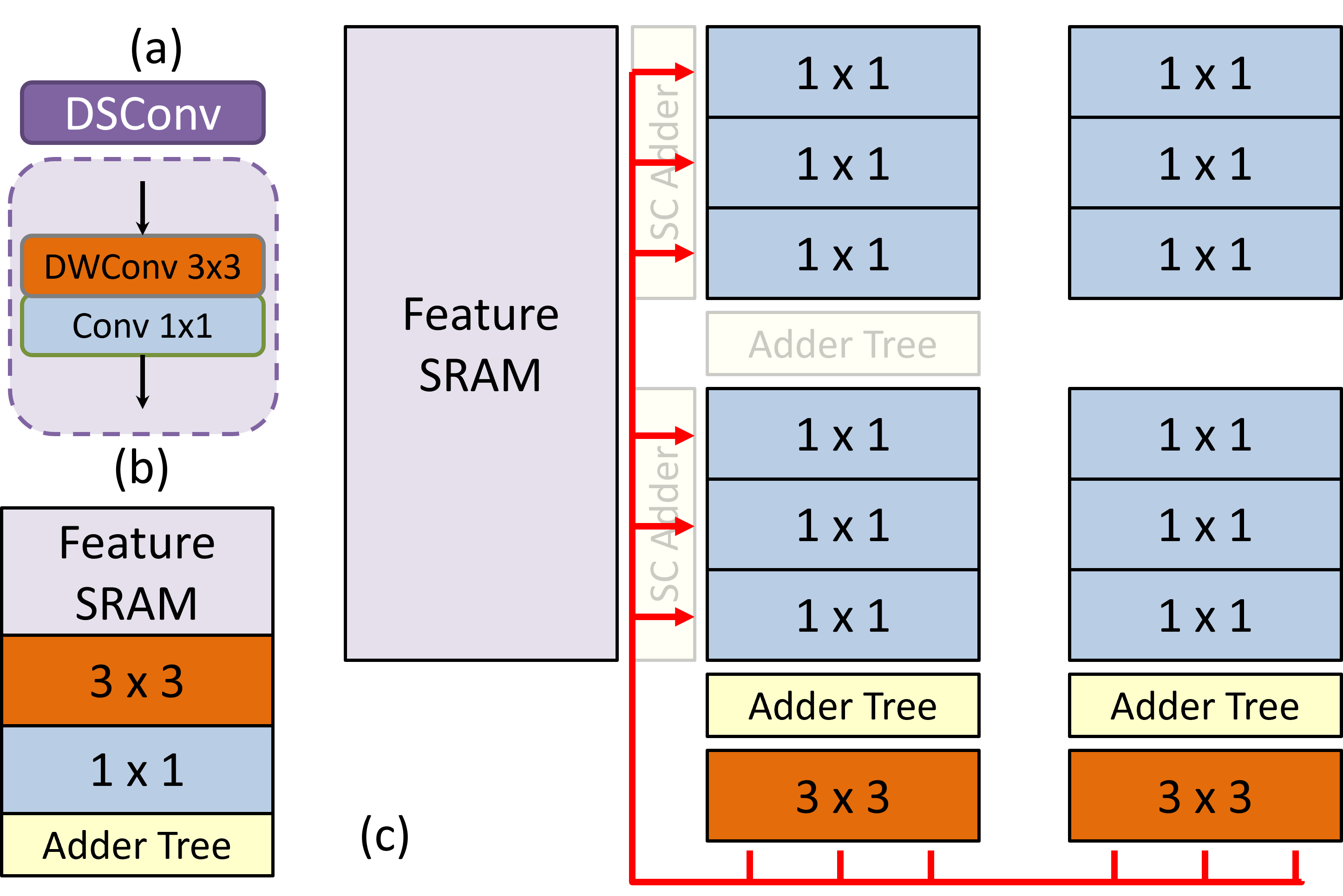}
        \caption{The dataflow of the DSConv in the C54 model. Best viewed in colors.}
        \label{HW_CLM_DSConv}
    \end{minipage}
    
\end{figure*}

\subsection{Overview of GLNPU}
Fig.~\ref{HW_overall_arch_v3} shows the proposed GLNPU that includes three SRAM buffers and group-of-layer PEs. This design dynamically maps groups of convolutional layers to PEs instead of the commonly used layer-by-layer processing method. This enables near-full hardware utilization and facilitates easy layer fusion processing, thus reducing feature SRAM accesses.

The input is a 32x32 patch according to Table~\ref{tab: patch size ablation study}. This size offers similar PSNR to larger patch sizes but incurs significantly smaller hardware area costs. In this design, to further minimize the DRAM bandwidth, we store the entire model weight and feature on-chip. This necessitates a 67KB weight SRAM, and three 69KB feature SRAMs with two for a ping-pong buffer and one for the shortcut. The boundary data between patches is stored in a 114KB boundary SRAM.

Each blue rectangle on the right-hand side of Fig.~\ref{HW_overall_arch_v3} represents a block consisting of 27x9 PEs. The detailed architectures of the 1×1 PE and 3×3 PE will be discussed in the later subsection.

\begin{table}
\centering
\caption{The ablation study of the patch size with $threshold_{1} = 8$ and $threshold_{2} = 40$ for the ×4 scale.}
\label{tab: patch size ablation study}
\begin{tblr}{
  width = \linewidth,
  colspec = {Q[198]Q[248]Q[198]Q[267]},
  cells = {c},
  row{1} = {font=\bfseries},
}
\toprule
Patch Size & {Test8K PSNR\\ (dB)} & {Line Buffer\\ (KB)} & {Feature SRAM\\ (KB)} \\
\midrule
16×16      & 34.63                & 2.36                 & 17                    \\
32×32      & 34.69                & 4.52                 & 69                    \\
48×48      & 34.70                & 6.68                 & 156                   \\
64×64      & 34.70                & 8.84                 & 276                   \\
\bottomrule
\end{tblr}
\end{table}

\subsection{Configurable group of layer mapping}
\label{section: Configurable group of layer mapping}

Due to the significant variations in the number of MACs among the bilinear, C27, and C54 models used for dynamic processing, achieving efficient processing in the PE array becomes challenging. The MACs of the C27 model amount to only 29.1\% of the MACs in the C54 model. Moreover, the bilinear model requires only 0.4\% of the MACs needed by the C54 model. To address this issue, we propose \textit{configurable group of layer mapping}, enabling high PE utilization when employing these three models in the same PE array.

PE arrays are allocated to process 54 channels of 1×1 point-wise convolution and 54 channels of 3×3 depth-wise convolution based on the model configuration and tradeoff of hardware cost and throughput requirement. With this configuration, we divide our ESSR model into three components: BSConv fusion, adding a shortcut with 1×1 convolution, and DSConv fusion. 

Fig.~\ref{HW_CLM_BSConv} to \ref{HW_CLM_DSConv} show the dataflow to map C54. Fig.~\ref{HW_CLM_SFB} shows the dataflow for C27. In these figures, (a) is the desired function, and (b) is the detailed function sequence with its corresponding physical dataflow in (c).

\subsection{C54 PE mapping}

The dataflow for the BSConv fusion component is depicted in Fig.~\ref{HW_CLM_BSConv} that executes 1x1 and 3x3 layers concurrently to save feature SRAM access by 43\%. Initially, the feature is loaded from the feature SRAM into the 1×1 PE array. It then passes through an adder tree to accumulate partial sums. Subsequently, the feature is fed into the 3×3 PE array to complete the BSConv operation. Finally, the resulting feature is stored back into the feature SRAM. 

Following above, Fig.~\ref{HW_CLM_SC1x1} shows the dataflow of adding a shortcut with 1×1 convolution. The feature and the shortcut are first added together in the shortcut adder. The combined feature is then entered into the 1×1 PE array. After passing through the adder tree to accumulate partial sums, the result is stored back into the feature SRAM.

Finally, the dataflow for the DSConv fusion component, as shown in Fig.~\ref{HW_CLM_DSConv}, involves directing the feature into the 3×3 PE array, followed by entering it into the 1×1 PE array. Once again, the adder tree is employed to accumulate partial sums. Since this layer represents the last convolution layer of the entire model, the final result undergoes boundary processing before being outputted to DRAM.

\subsection{C27 PE mapping}
\label{section: C27 PE mapping}
\subsubsection{SFB: Structure-Friendly Fusion Block}
\label{section: SFB}

\begin{figure*}[!t]
    \centering
    
    \begin{minipage}{0.48\linewidth}
        \centering
        \includegraphics[width=\linewidth,keepaspectratio=true]{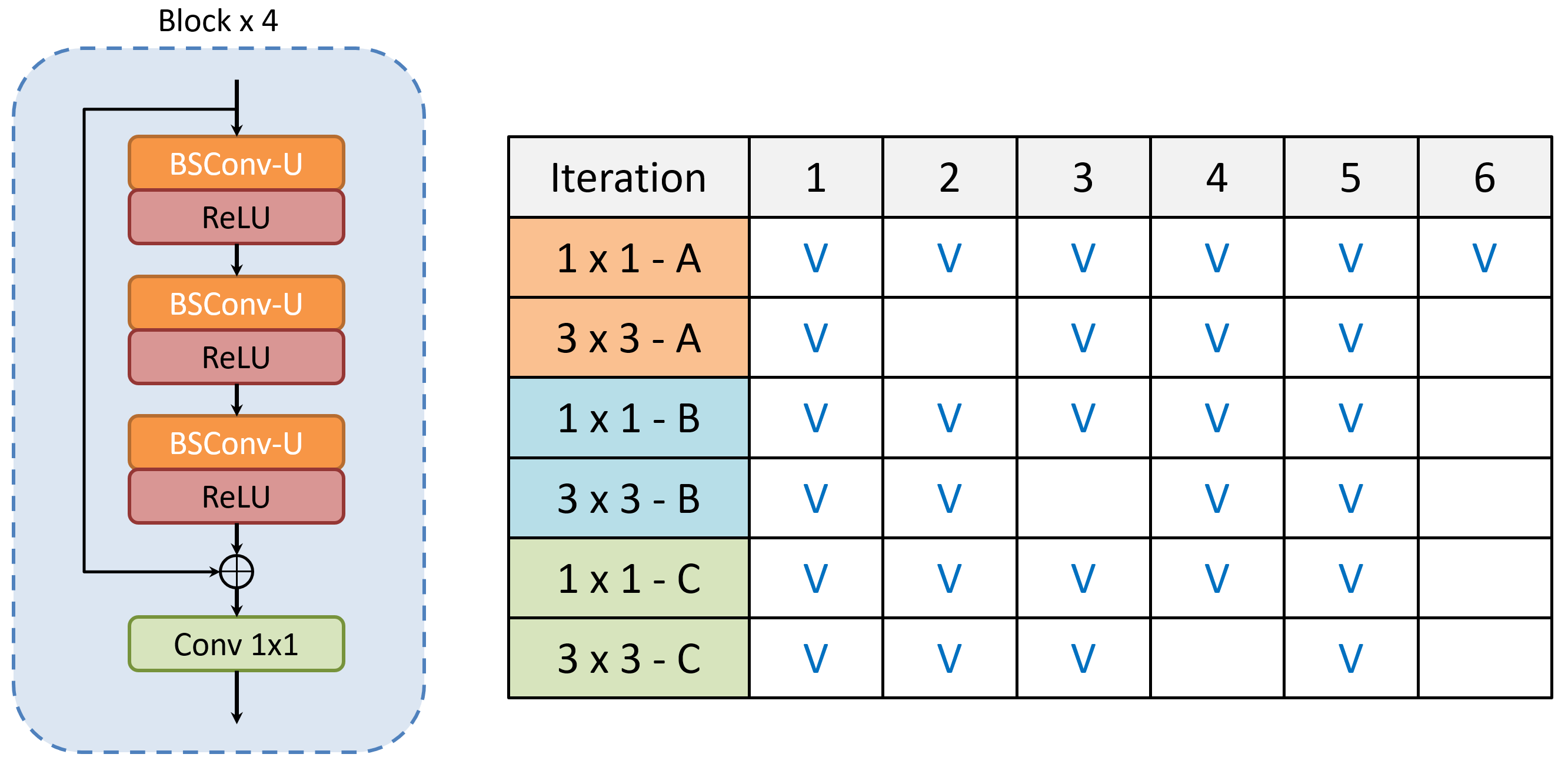}
        \caption{The BSConv in the RLFB and its scheduling.}
        \label{block×4 each iter}
    \end{minipage}
    \hfill
    \begin{minipage}{0.48\linewidth}
        \centering
        \includegraphics[width=\linewidth,keepaspectratio=true]{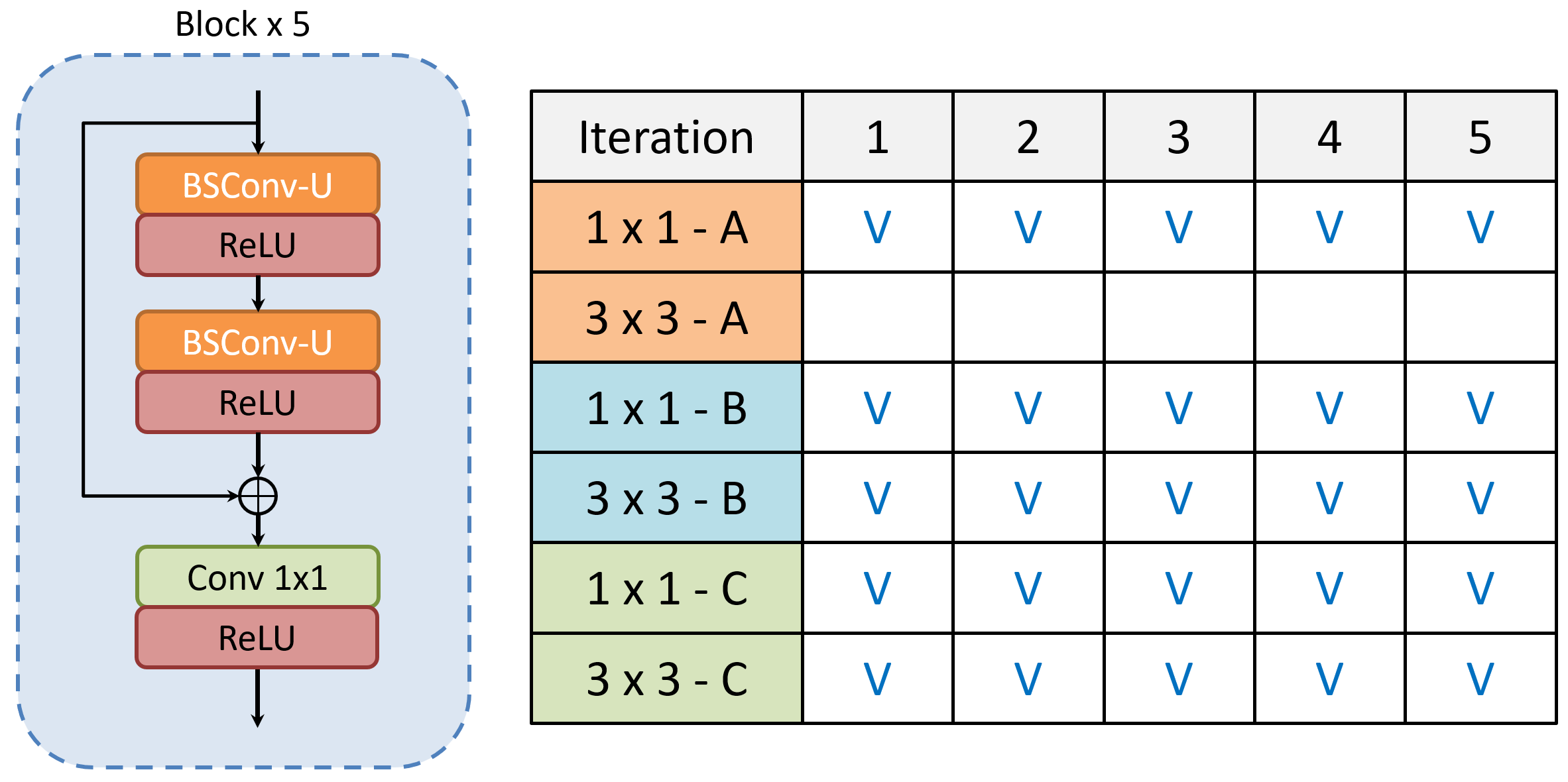}
        \caption{Proposed SFB and its scheduling.}
        \label{SFB each iter}
    \end{minipage}
    
\end{figure*}

In the original RLFB of RLFN, the structure is composed of three convolutions, succeeded by the integration of shortcuts and a 1×1 point-wise convolution. Given our PE array's capability, it can accommodate one 1×1 point-wise convolution and a singular 3×3 depth-wise convolution during each iteration for the C54 model. For the C27 model, the identical PE allocation can process four 1×1 point-wise convolutions and a pair of 3×3 depth-wise convolutions within the same iteration span. This infers that approximately one and a third iterations are mandated to execute a singular BSConv variant of RLFB. Implementing this in hardware would need six iterations to process four such blocks, leading to complex and varying PE controls across iterations. Fig.~\ref{block×4 each iter} provides a visual representation of its hardware scheduling across iterations, highlighting its irregularity and suboptimal hardware utilization.

To alleviate these complexities and uphold efficiency, we introduce SFBs. An SFB is architecturally defined by a pair of BSConv layers, which are succeeded by the incorporation of shortcuts and a 1×1 point-wise convolution. Furthermore, we integrate a ReLU activation at the end of the SFB, enabling zero gating in the subsequent BSConv layer. 
Fig.~\ref{SFB each iter} shows its hardware scheduling across various iterations. The configuration is tailored to process the complete block within a single iteration, thereby maximizing hardware utilization.

Taking advantage of the above modifications, to maintain the performance of the original model, we use five SFBs in our design, as detailed in the ablation study encapsulated in Table~\ref{tab:SFB ablation study}. This configuration requires five iterations, which saves one iteration compared to the four-RLFB version.

\begin{table}[tb]
\centering
\caption{The ablation study of the number of the SFB. \textit{w/o Bias} is without the bias of convolution.}
\label{tab:SFB ablation study}
\begin{tblr}{
  width = \linewidth,
  colspec = {lcc},
  row{1} = {font=\bfseries},
}
\toprule
\#SFBs    & Params(K) & Set5 PSNR(dB) \\
\midrule
4         & 43.9      & 31.46         \\
5         & 53.9      & 31.51         \\
5-w/o Bias & 53.6      & 31.46         \\
6         & 63.9      & 31.49         \\
\bottomrule
\end{tblr}
\end{table}

\subsubsection{SFB mapping}
\begin{figure}[tb]
    \centering
    \includegraphics[height=!,width=1\linewidth,keepaspectratio=true]{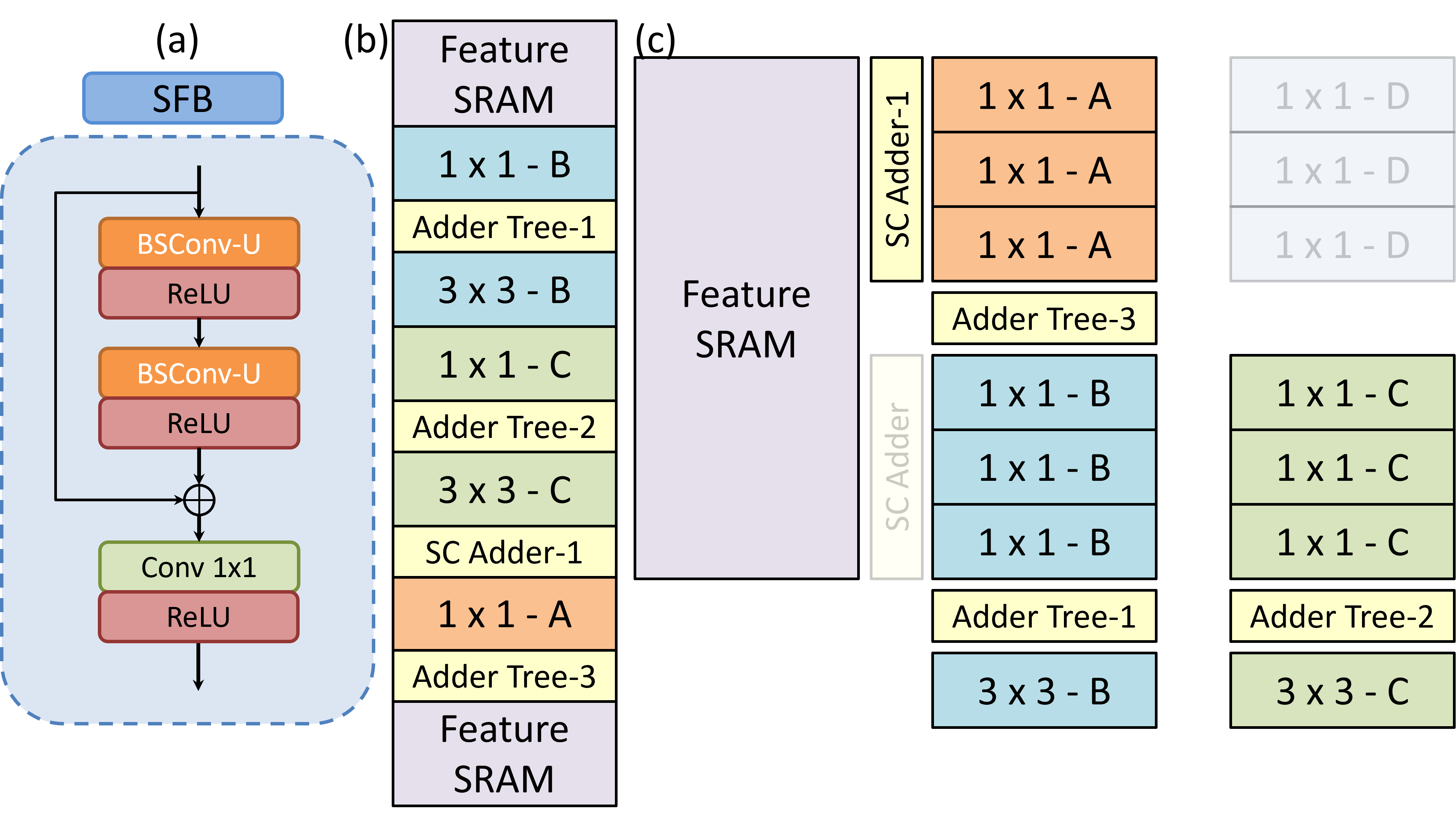}
    \caption {The dataflow of the SFB in the C27 model. Best viewed in colors.}
    \label{HW_CLM_SFB}
\end{figure}

With the above modifications, our PE array can handle the entire SFB block of the C27 model. Fig.~\ref{HW_CLM_SFB} shows its dataflow that executes the whole SFB layers concurrently to save 79
\% of feature SRAM access. In each iteration, the feature is initially loaded into the 1×1-B PE array. After the adder tree sums up the three partial sums, the feature is directed to the 3×3-B PE array, completing the first BSConv operation. The second BSConv operation is conducted using the 1×1-C PE array and the 3×3-C PE array, following the same flow as the first BSConv. Finally, the feature and the shortcut are combined in the shortcut adder. The resultant combined feature is then inputted into the 1×1-A PE array. After passing through the adder tree to accumulate partial sums, the final result is stored back into the feature SRAM.

\subsection{Blininear interpolation mapping}
\label{section: Blininear interpolation PE mapping}
The bilinear interpolation is mapped to the 3×3 PE arrays that load its corresponding source pixels from the feature SRAM and undergo boundary processing and store output to the DRAM.

\subsection{Detailed architecture of the 1×1 and 3x3 blocks}
\label{section: 1x1 PE Detailed architecture}

\begin{figure*}[!t]
    \centering
    
    \begin{minipage}{0.48\linewidth}
        \centering
        \includegraphics[width=0.8\linewidth,keepaspectratio=true]{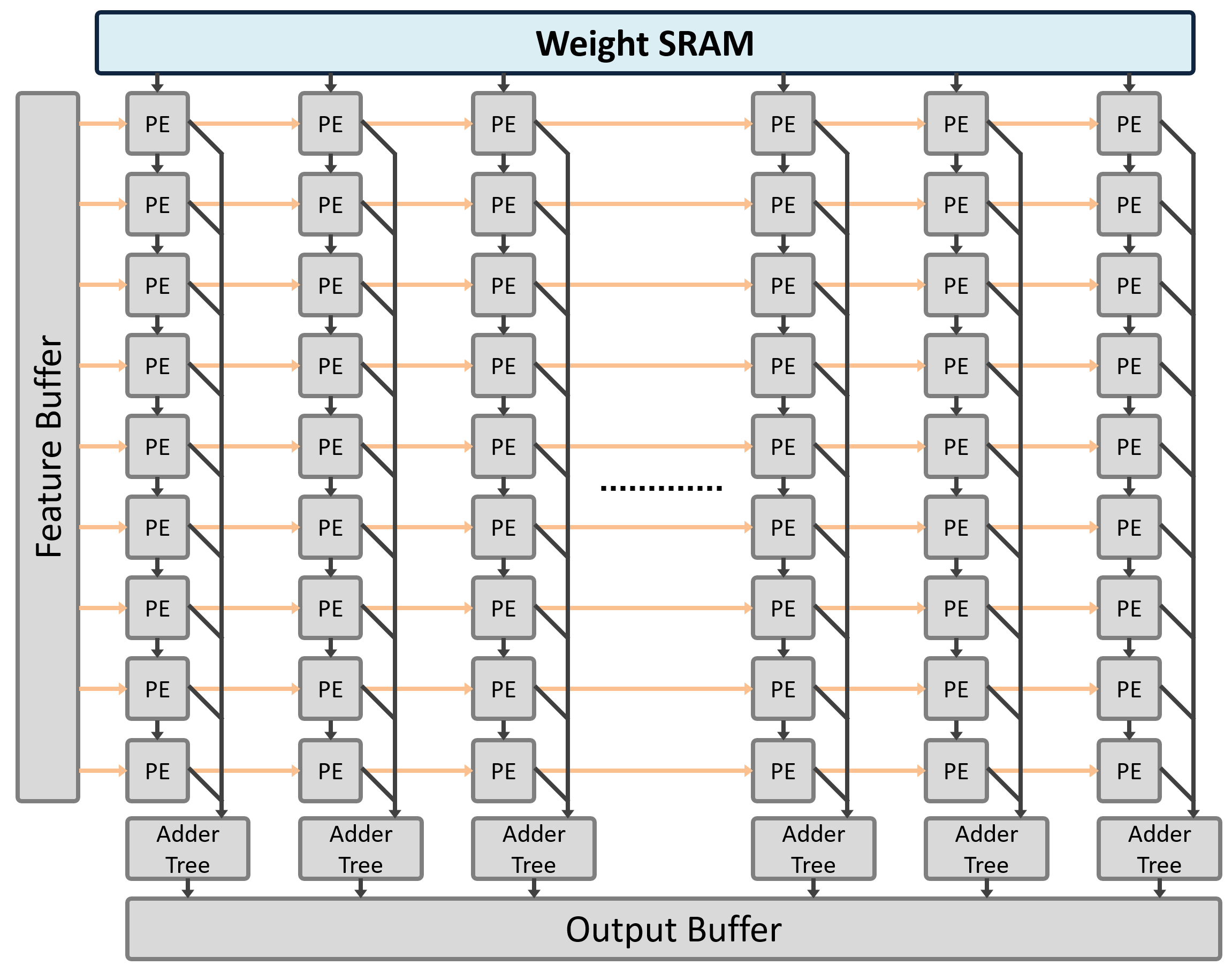}
        \caption{The details of the 1×1 PE block.}
        \label{HW_1x1_detail}
    \end{minipage}
    \hfill
    \begin{minipage}{0.48\linewidth}
        \centering
        \includegraphics[width=0.8\linewidth,keepaspectratio=true]{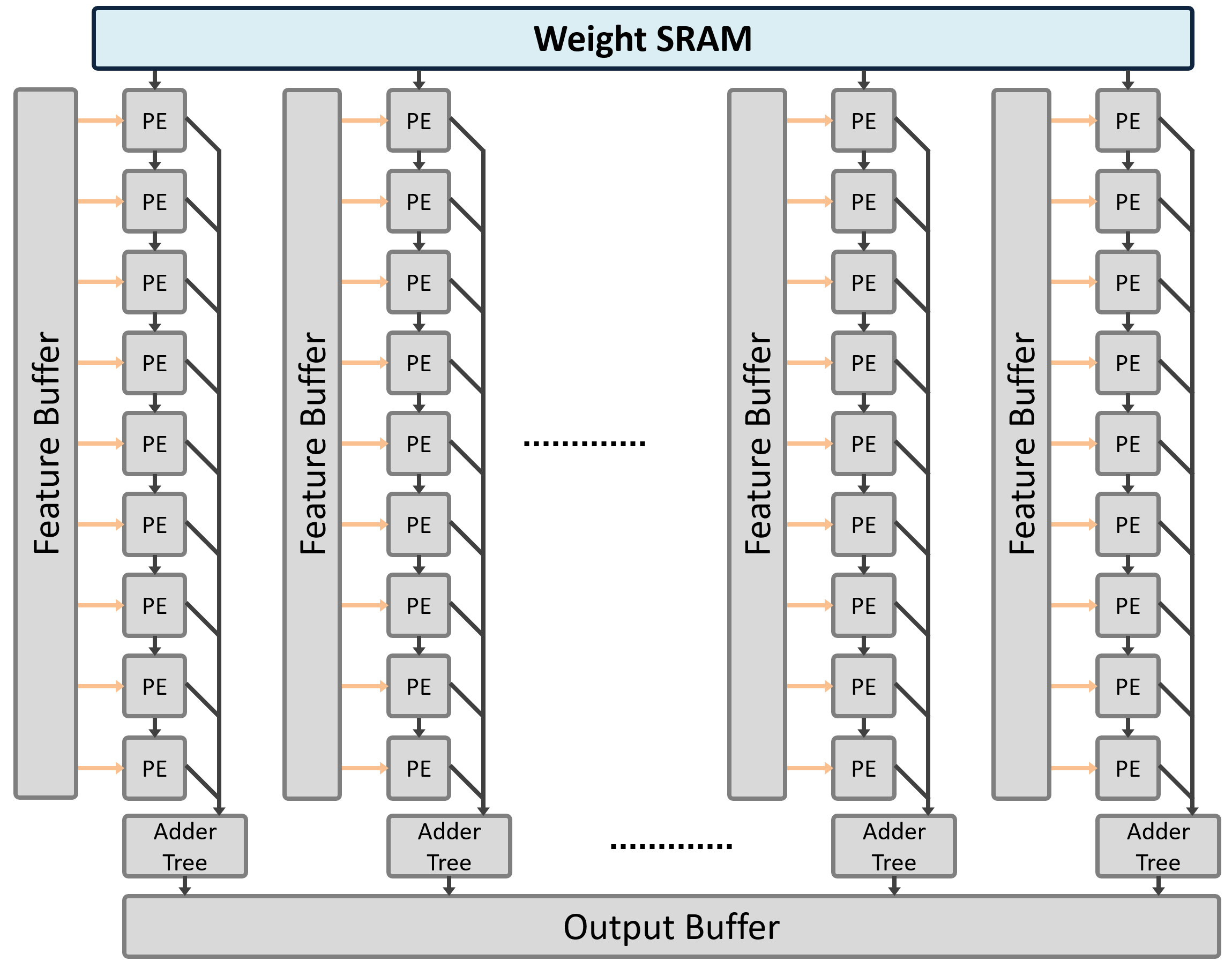}
        \caption{The details of the 3×3 PE block.}
        \label{HW_3x3_detail}
    \end{minipage}
    
\end{figure*}

Fig.~\ref{HW_1x1_detail} shows the detailed design of a 1x1 block that contains 27x9 PEs. Weights are preloaded to each PE at the beginning in a pipeline way. This occurs when the last 1×1 convolution is finished while the 3×3 convolution is still processing, which can save long loading cycles. Then the weights remain stationary during computing to save power. The feature inputs are broadcast to all PEs. Each PE then multiplies the weights with the feature. In each cycle, the adder tree sums the nine products from each column of PEs and saves the partial sum in the output buffer. The 3x3 block shown in Fig.~\ref{HW_3x3_detail} also contains 27x9 PEs and uses a similar operation flow.

\subsection{Quantization}
\label{sec: Quantization}
The proposed model is quantized with PAMS \cite{PAMS}. PAMS quantizes both weights and activations to 8-bit fixed-point, but excludes the first and last convolutions to avoid performance impact. However, this approach requires hardware for both floating-point (FP) PE and fixed-point (FXP) PE, thereby increasing complexity and area. Thus, we quantize the whole ESSR with PAMS but with 10-bit FXP, resulting in a 0.03dB decrease in PSNR.

\subsection{Boundary Processing}
\label{section: Boundary Processing}
For patch-based processing, a critical challenge is managing the boundary data interspersing the patches. Ideally, lossless boundary processing can be achieved by preserving the boundary feature and recalculating it in adjacent patches, essentially converting block convolution back to its conventional form. Various storage solutions exist for these boundary data: they can be retained in SRAM as discussed in \cite{ACNPU}, stored in DRAM, or managed using a hybrid approach as highlighted in \cite{SRNPU2022}. Nonetheless, these methods can be resource-intensive due to the voluminous input and the complexity of models.

Another strategy involves computing the patch with overlapping boundaries and subsequently averaging the results from these overlaps. While this approach is straightforward, it leads to significant computational overhead. A more cost-efficient alternative involves non-overlapped patch convolution, coupled with post-processing techniques designed to rehabilitate the boundary pixels. This method, despite its efficiency, often results in pronounced quality degradation and introduces artifacts.

A comprehensive cost analysis of various boundary processing solutions is provided in Table~\ref{boundary processing solutions}. To replicate the performance observed with software-based whole image input, one would either need to allocate 9.02GB of DRAM bandwidth or 1672KB for boundary SRAM. Unfortunately, both options require substantial hardware resources, increasing implementation costs. The results derived from the naive interpolation strategy are unsatisfactory.

Consequently, in the pursuit of a balanced solution, we opt for the 'overlap and average' technique. As illustrated in Table~\ref{compare overlap and average}, an overlap of 8 pixels can yield nearly the same performance as an overlap of 16 pixels, with only a 14\% increase in MACs. Furthermore, the boundary SRAM footprint of this method proves to be more acceptable when compared to the 'save SRAM and recompute' approach. Taking into account these factors, we chose the overlap and average method with an 8-pixel overlap, which incurs a reasonable hardware cost and results in only a 0.03dB PSNR drop on Set14.

\begin{table}
\centering
\caption{The comparison of boundary processing solutions. Urban: Urban100, Manga: Manga109}
\label{boundary processing solutions}
\fontsize{8}{10}\selectfont
\begin{tblr}{
  width = \linewidth,
  colspec = {lcccccl},
  row{1} = {font=\bfseries},
}
\toprule
Method       & {SRAM\\(KB)} & {BW\\(GB/s)} & Set5  & Urban  & Manga \\
\midrule
Interpol.    & 13           & 3.17         & 31.41 & 25.14 & 29.02    \\
DRAM+Recomp. & 48           & 9.02         & 31.51 & 25.28 & 29.34    \\
SRAM+Recomp  & 1672         & 3.17         & 31.51 & 25.28 & 29.34    \\
Overlap+Avg. & 114          & 3.61         & 31.46 & 25.23 & 29.23   \\
\bottomrule
\end{tblr}
\end{table}

\begin{table}[tb]
\centering
\caption{The effect of the overlapped pixels. Urban: Urban100, Manga: Manga109}
\label{compare overlap and average}
\begin{tblr}{
  width = \linewidth,
  colspec = {lcccccc},
  row{1} = {font=\bfseries},
}
\toprule
Overlap & {SRAM \\ (KB)} & {BW\\ (GB/s)} & MACs  & Set5  & Urban  & Manga \\
\midrule
16      & 243       & 4.14      & 131\% & 31.46 & 25.24    & 29.23    \\
12      & 176       & 3.86      & 122\% & 31.45 & 25.24    & 23.24    \\
8       & 114       & 3.61      & 114\% & 31.46 & 25.23    & 29.23    \\
4       & 55        & 3.38      & 107\% & 31.44 & 25.21    & 29.19    \\
0       & 0         & 3.17      & 100\% & 31.41 & 25.14    & 29.02    \\
\bottomrule
\end{tblr}
\end{table}

Following our detailed analysis, we adopt a method termed as the \textit{slim overlap block convolution and thick overlap boundary processing}. In this approach, we introduce an overlap of 2 pixels within the LR patches during the block convolution phase. Once we execute a ×4 scale upsampling, this results in patches that feature an overlap of 8 pixels.

\section{Experimental Results}
\subsection{Setup}
The proposed model is trained on Flickr2K \cite{Flicker2K} and DIV2K \cite{DIV2K}.
Evaluation metrics include the commonly used peak signal-to-noise ratio (PSNR) and the structural similarity (SSIM) on the Y-channel as well as the perceptual one, LPIPS, on the RGB channels.

The training for the PSNR-oriented model adopts the L1 loss with the batch size set to 256, the Lamb optimizer, and the initial learning rate set to 3e-3 with cosine learning rate decay. The model is trained for 200,000 iterations. To improve perceptual quality, we also apply perceptual-oriented training. For the perceptual-oriented model, we load the trained PSNR-oriented model as an initial model, set the batch size to 16 and use Adam optimizer with the initial learning rates set to 1e-4 with MultiStepLR. To optimize this model, we employ L1 loss, artifact loss, perceptual loss, and adversarial loss from \cite{LDL} for a total of 300,000 iterations. The weighted combination of each loss is set to 0.01, 1, 1, and 0.005, respectively.

The training methodology for the supernet that includes C27 and C54 follows that proposed by ARM~\cite{ARM}. Subnet selection is chosen based on the probability proportional to the MACs of each subnet. In each iteration, only one subnet is chosen and the parameters of that selected subnet are updated. Both types of models use an exponential moving average (EMA) with a decay of 0.999 to update weights and improve training stability and overall performance. Our model implementation is based on PyTorch and runs on a single NVIDIA DGX Station A100.

\subsection{Simulation Results}

\begin{figure}[tb]
    \centering
    \includegraphics[height=!,width=1\linewidth,keepaspectratio=true]{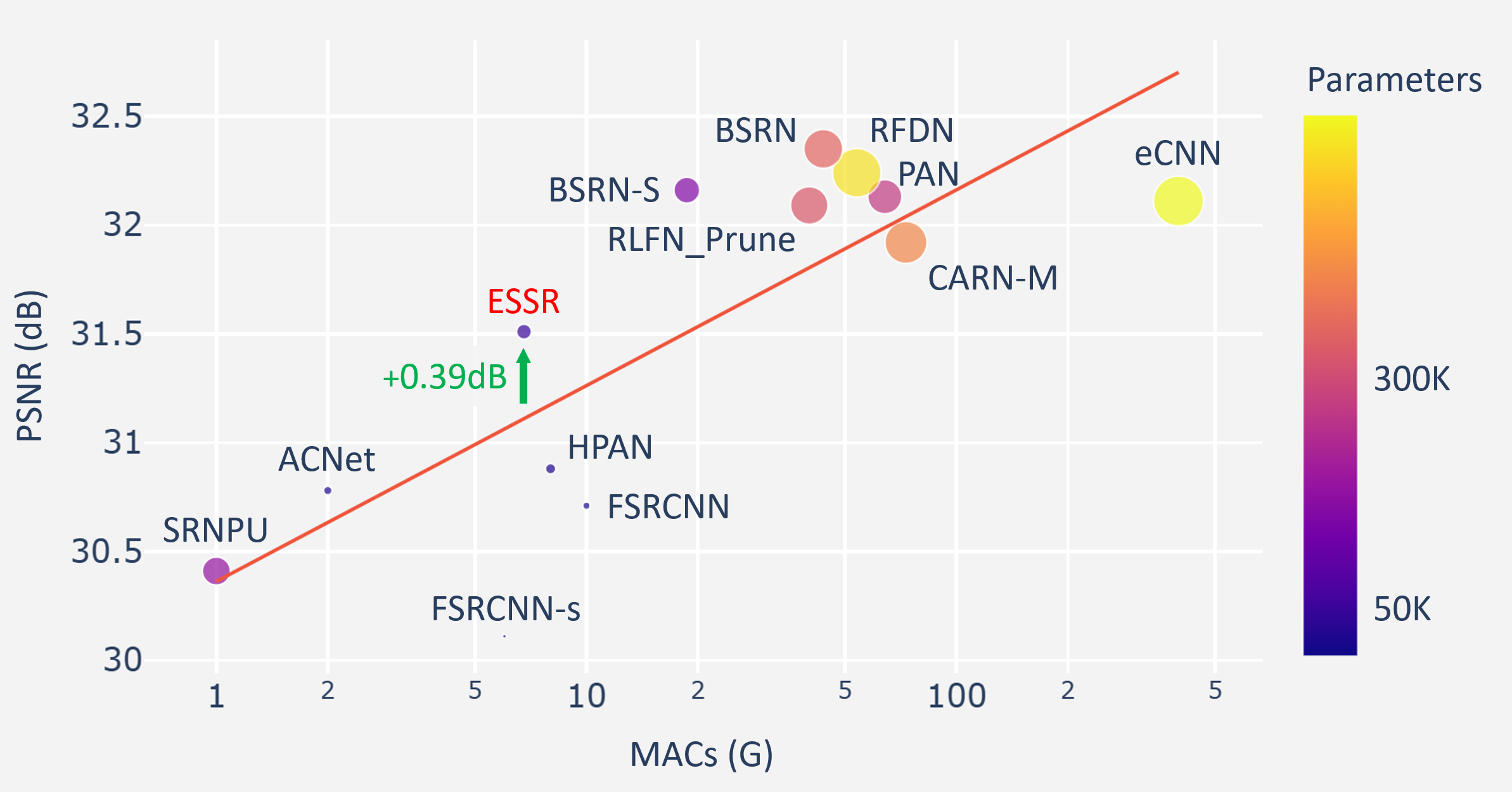}
    \caption {Performance and model complexity comparison on the Set5 dataset for the ×4 scale.}
    \label{MACs_PSNR_Parameter_Set5_×4_1920x1080output_wT_wA}
\end{figure}

Fig.~\ref{MACs_PSNR_Parameter_Set5_×4_1920x1080output_wT_wA} shows a comparison of PSNR, MACs, and parameters. Our model demonstrates a notable improvement of 0.39dB above the trendline in terms of PSNR. Moreover, it stands out as one of the most outstanding models within the less than 10G MACs region.

\subsubsection{Comparison with other SR networks}
Table~\ref{tab: ×2 scale quantitative results} and Table~\ref{tab: ×4 scale quantitative results} show the quantitative performance comparison, sorted by Set5 PSNR. Remarkably, ESSR outperforms other state-of-the-art models in terms of PSNR, specifically in the under 100K parameter region, while employing nearly the fewest MACs. Furthermore, ESSR achieves comparable or even superior performance compared to LapSRN \cite{LapSRN}, with the advantage of using 760K fewer parameters. This signifies the effectiveness and efficiency of our proposed ESSR model, highlighting its competitive advantage over existing approaches.

Compared to other SR accelerators like SRNPU, ACNet, and eCNN, ESSR shows the best performance in the ×2 scale category and the second best performance in the ×4 scale category. However, it is worth noting that eCNN achieves the highest PSNR in the ×4 scale category, but has 10 times more parameters and more than 13 times more MACs compared to ESSR. 

When comparing ESSR with the other dynamic processing method like SRNPU, ESSR has 130K fewer parameters. This reduction is attributed to the absence of an additional prediction network in ESSR and the sharing of parameters among its subnets. Regarding MACs and PSNR, ESSR has slightly higher MACs but with 1dB gain. This indicates that despite the slightly higher computational cost, ESSR provides superior image quality enhancement, making it a favorable choice in terms of performance.

\begin{table}
\centering
\caption{The ×2 scale quantitative results of some PSNR-oriented light weight models. The MACs is calculated with a 1920x1080 GT image. $^*$ This model has its own hardware design. The units for size and MAC are K and G, respectively. Urban: Urban100, Manga: Manga109}
\label{tab: ×2 scale quantitative results}
\fontsize{8}{10}\selectfont
\begin{tabularx}{\columnwidth}{c|c|c|X|X|X|X|X}
\toprule
\textbf{Model}&\textbf{Size}&\textbf{MAC}&\textbf{Set5}&\textbf{Set14}&\textbf{B100}&\textbf{Urban}&\textbf{Manga}\\ 
\hline
Bicubic             & 1             & -            & 33.66          & 30.24          & 29.56          & 26.88          & 30.8            \\
FSRCNN-s            & 4             & 6            & 36.57          & 32.28          & 31.23          & -              & -               \\
SRCNN               & 57            & 119          & 36.66          & 32.45          & 31.36          & 29.50          & 35.60           \\
FSRCNN              & 13            & 14           & 37.00          & 32.63          & 31.53          & 29.88          & 36.67           \\
SRNPU$^*$           & 183           & 7            & 37.06          & 32.62          & 31.47          & -              & -               \\
ACNet$^*$           & 17            & 9            & 37.34          & 32.78          & 31.64          & 30.21          & 36.88           \\
HPAN                & 26            & 17           & 37.38          & 32.91          & 31.69          & 30.29          & 37.00           \\
LapSRN              & 813                    & 67                & 37.52          & 33.08          & 31.80          & 30.41             & -                 \\
CARN-M              & 412           & 182          & 37.53          & 33.26          & 31.92          & 31.23          & -               \\
eCNN$^*$            & 202           & 106          & 37.62          & 33.17          & 31.93          & 30.60          & -               \\
\textbf{ESSR} & \textbf{51}   & \textbf{26}  & \textbf{37.64} & \textbf{33.18} & \textbf{31.90} & \textbf{30.92} & \textbf{37.89}  \\
PAN                 & 261           & 160          & 38.00          & 33.59          & 32.18          & 32.01          & 38.70           \\
RFDN                & 534           & 284          & 38.05          & 33.68          & 32.16          & 32.12          & 38.88           \\
RLFN                & 527           & 260          & 38.07          & 33.72          & 32.22          & 32.33          & -               \\
BSRN                & 332           & 164          & 38.10          & 33.74          & 32.24          & 32.34          & 39.14           \\
\bottomrule
\end{tabularx}
\end{table}

\begin{table}
\centering
\caption{The ×4 scale quantitative results of some PSNR-oriented light weight models. The MACs is calculated with a 1920x1080 GT image. $^*$ This model has its own hardware design. The units for size and MAC are K and G, respectively. Urban: Urban100, Manga: Manga109}
\label{tab: ×4 scale quantitative results}
\fontsize{8}{10}\selectfont
\begin{tabularx}{\columnwidth}{c|c|c|X|X|X|X|X}
\toprule
\textbf{Model}&\textbf{Size}&\textbf{MAC}&\textbf{Set5}&\textbf{Set14}&\textbf{B100}&\textbf{Urban}&\textbf{Manga}\\ 
\hline
Bicubic & - & - & 28.42 & 26.00 & 25.96 & 23.14 & 24.89 \\
FSRCNN-s & 4 & 6 & 30.11 & 27.19 & 26.84 & - & - \\
SRNPU$^*$ & 183 & 1 & 30.41 & 27.37 & 26.86 & - & - \\
SRCNN & 57 & 119 & 30.48 & 27.49 & 26.90 & 24.52 & 27.66 \\
FSRCNN & 13 & 10 & 30.71 & 27.59 & 26.98 & 24.62 & 27.90 \\
ACNet$^*$ & 18 & 2 & 30.78 & 27.62 & 27.00 & 24.63 & 27.82 \\
HPAN & 26 & 8 & 30.88 & 27.68 & 27.03 & 24.69 & 28.04 \\
\textbf{ESSR} & \textbf{53} & \textbf{7} & \textbf{31.51} & \textbf{28.22} & \textbf{27.31} & \textbf{25.28} & \textbf{29.34} \\
LapSRN              & 813                & 336              & 31.54          & 28.19          & 27.32          & 25.21             & 29.09             \\
CARN-M & 412 & 73 & 31.92 & 28.42 & 27.44 & 25.62 & - \\
eCNN$^*$ & 620 & 96 & 31.97 & 28.61 & 27.59 & 26.11 & - \\
PAN & 272 & 64 & 32.13 & 28.61 & 27.59 & 26.11 & 30.51 \\
RLFN & 543 & 67 & 32.24 & 28.62 & 27.60 & 26.17 & - \\
\bottomrule
\end{tabularx}
\end{table}

\begin{figure*}[!t]
    \centering
    
    \begin{minipage}{\textwidth}
        \centering
        \includegraphics[width=0.48\linewidth,keepaspectratio=true]{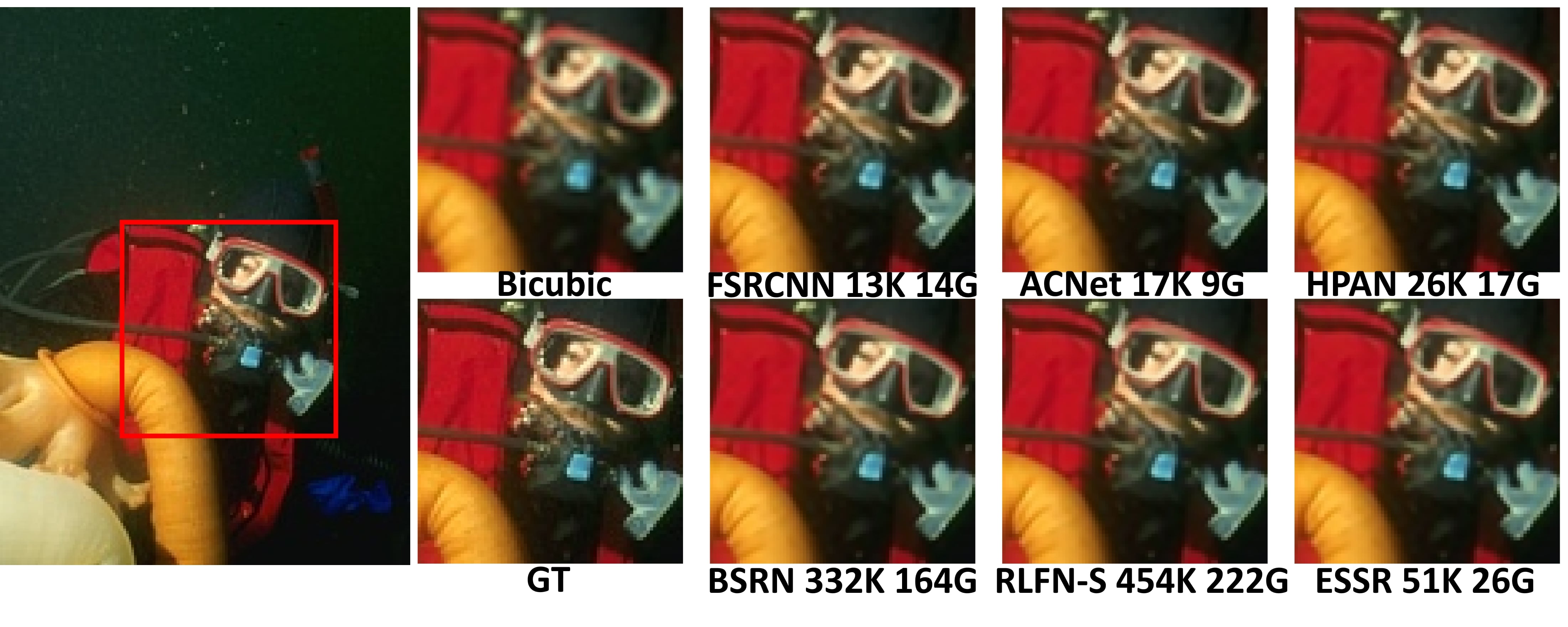}
        \hfill
        \includegraphics[width=0.48\linewidth,keepaspectratio=true]{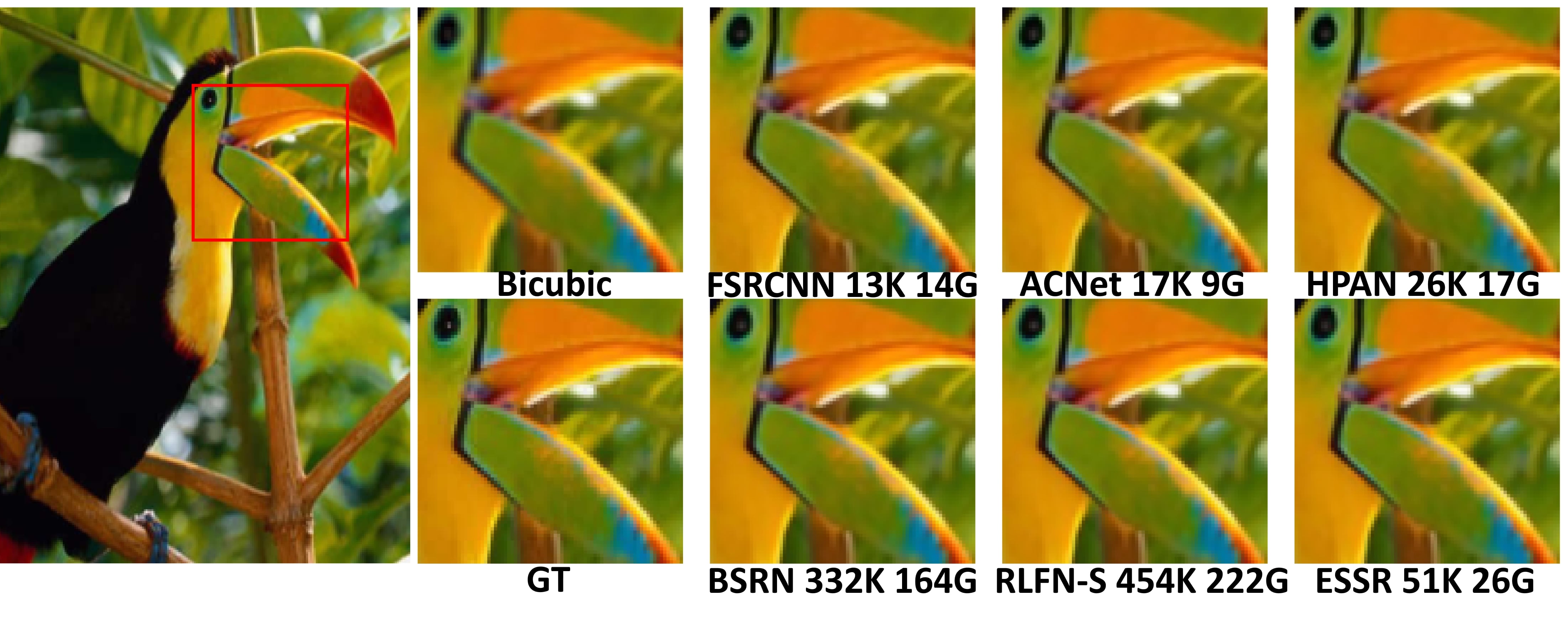}
        \caption{Visual comparison of the PSNR-oriented ESSR with the state-of-the-art methods. (Parameter/MACs). \textbf{(left)} $\times$2 B100 45096. \textbf{(right)} $\times$2 Set5 bird.}
        \label{SW_Image_Comparison_X2_123}
    \end{minipage}

    \begin{minipage}{\textwidth}
        \centering
        \includegraphics[width=0.48\linewidth,keepaspectratio=true]{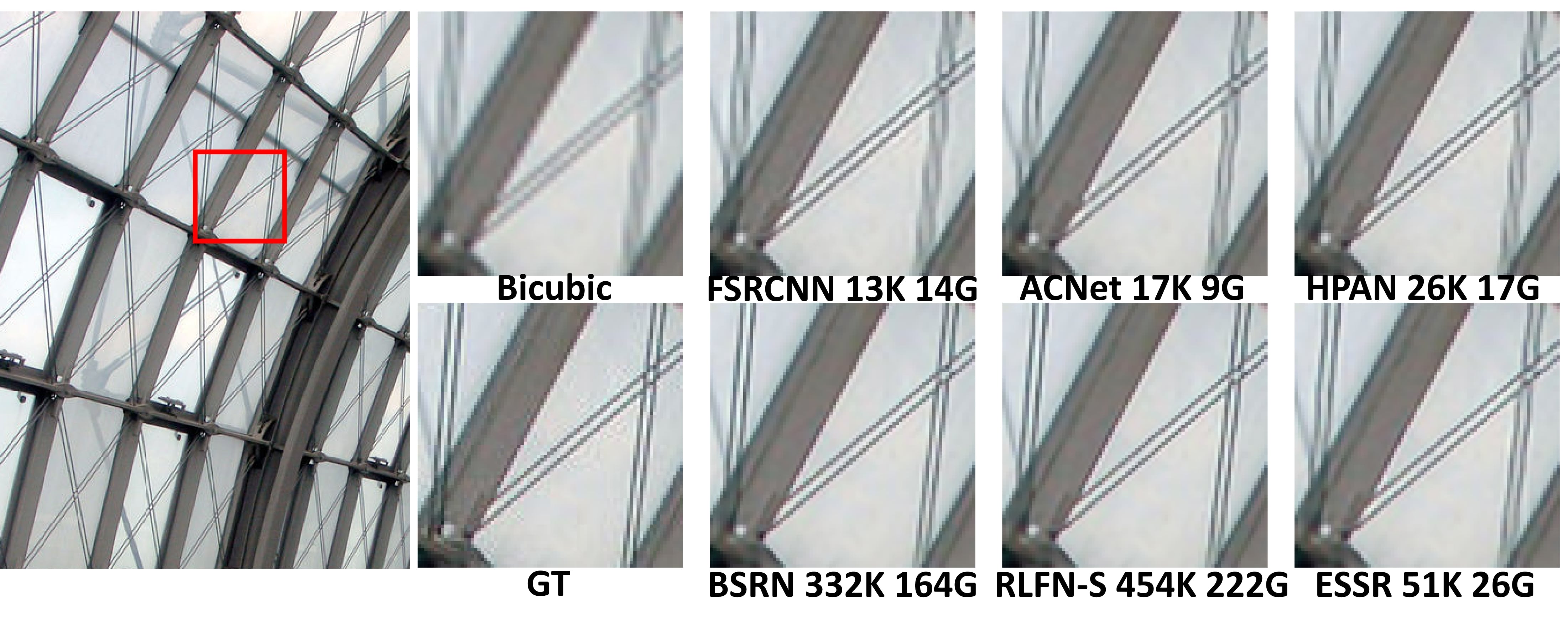}
        \hfill
        \includegraphics[width=0.48\linewidth,keepaspectratio=true]{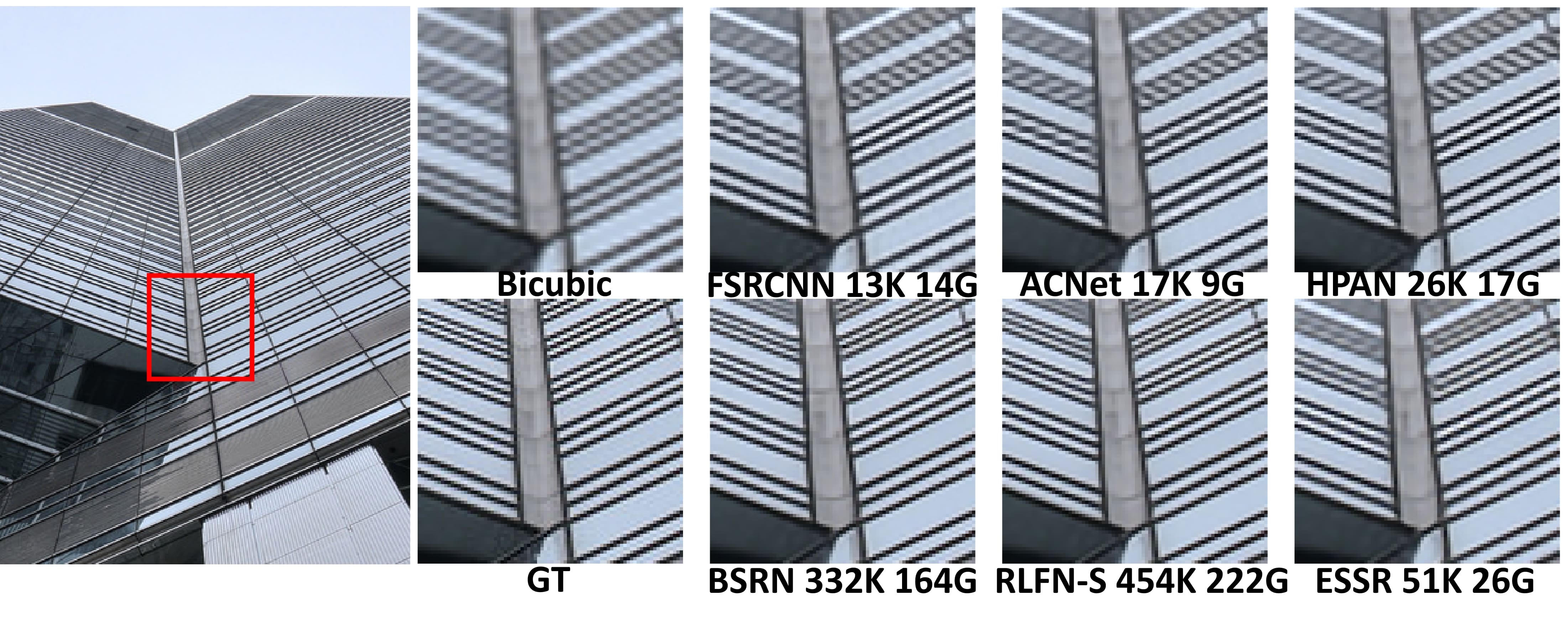}
        \caption{Visual comparison of the PSNR-oriented ESSR with the state-of-the-art methods. (Parameter/MACs). \textbf{(left)} $\times$2 Urban100 img002. \textbf{(right)} $\times$2 Urban100 img059.}
        \label{SW_Image_Comparison_X2_456}
    \end{minipage}

    \begin{minipage}{\textwidth}
        \centering
        \includegraphics[width=0.48\linewidth,keepaspectratio=true]{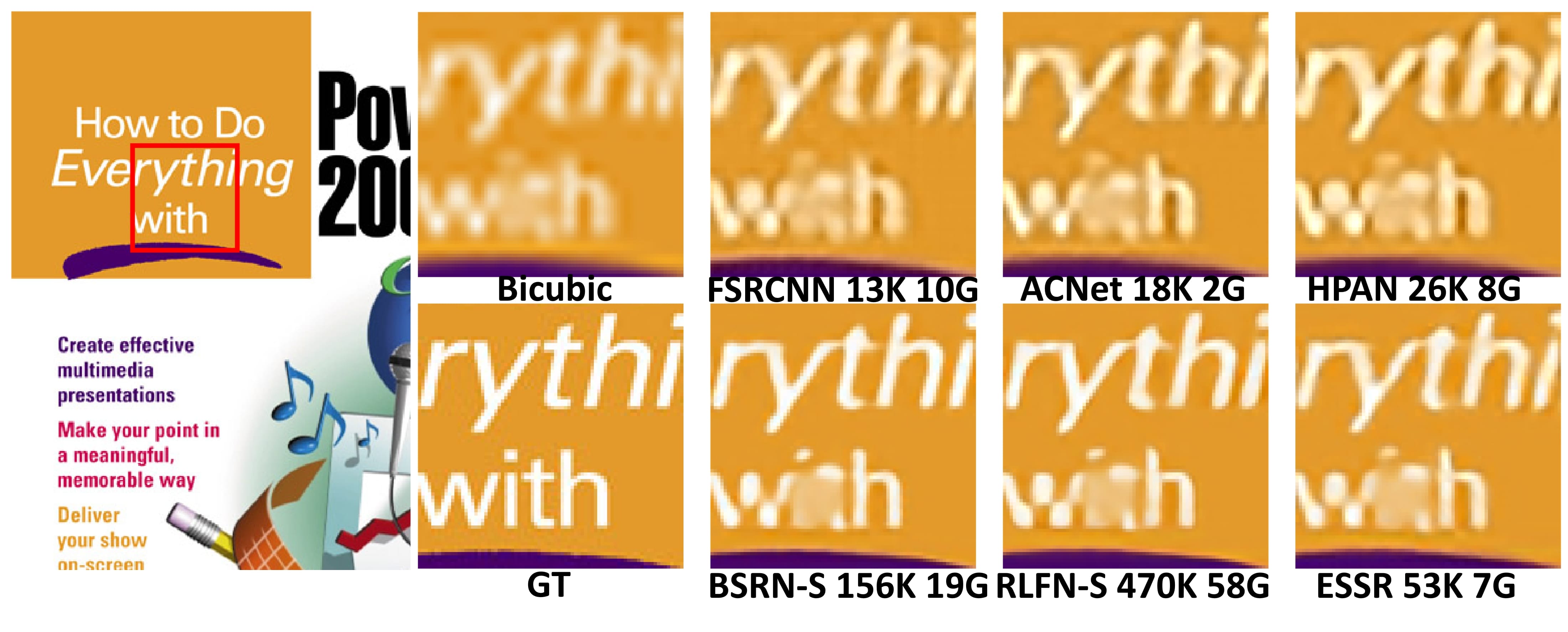}
        \hfill
        \includegraphics[width=0.48\linewidth,keepaspectratio=true]{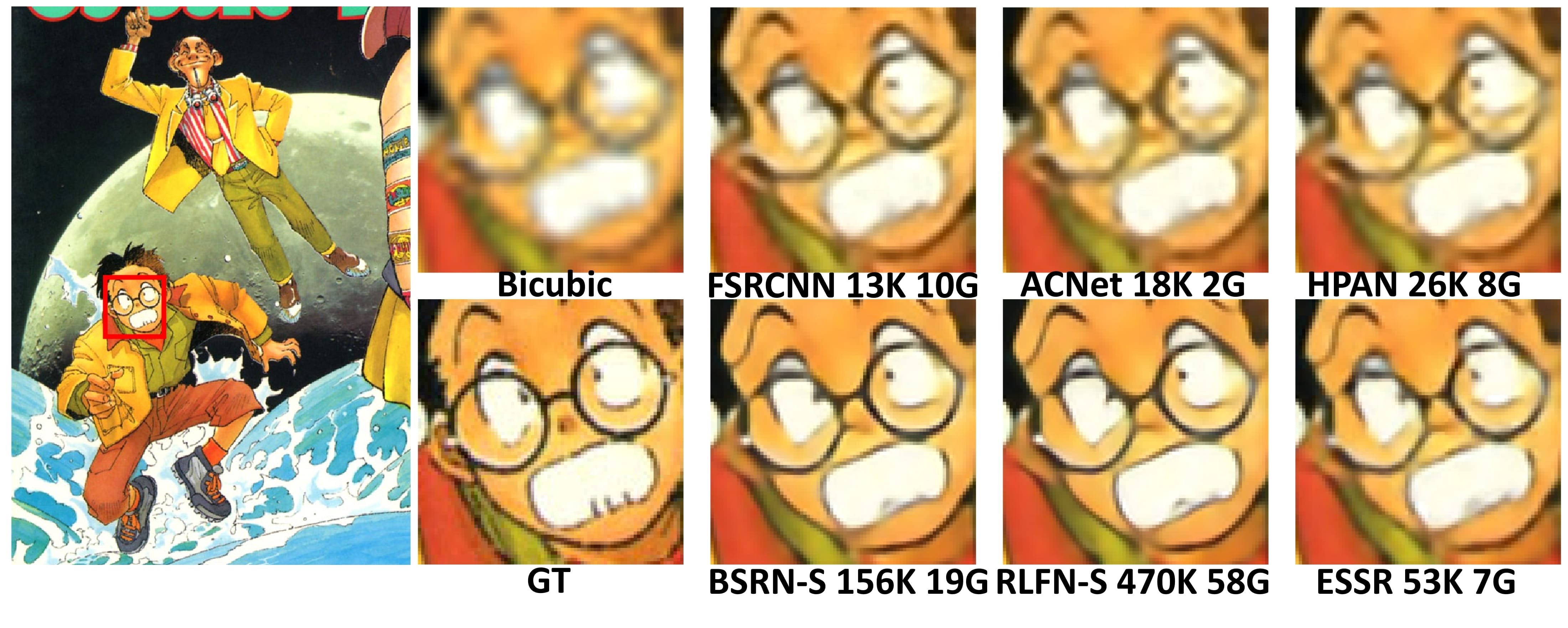}
        \caption{Visual comparison of the PSNR-oriented ESSR with the state-of-the-art methods. (Parameter/MACs). \textbf{(left)} $\times$4 Set14 ppt3. \textbf{(right)} $\times$4 Manga109 AppareKappore.}
        \label{SW_Image_Comparison_X4_123}
    \end{minipage}

    \begin{minipage}{\textwidth}
        \centering
        \includegraphics[width=0.48\linewidth,keepaspectratio=true]{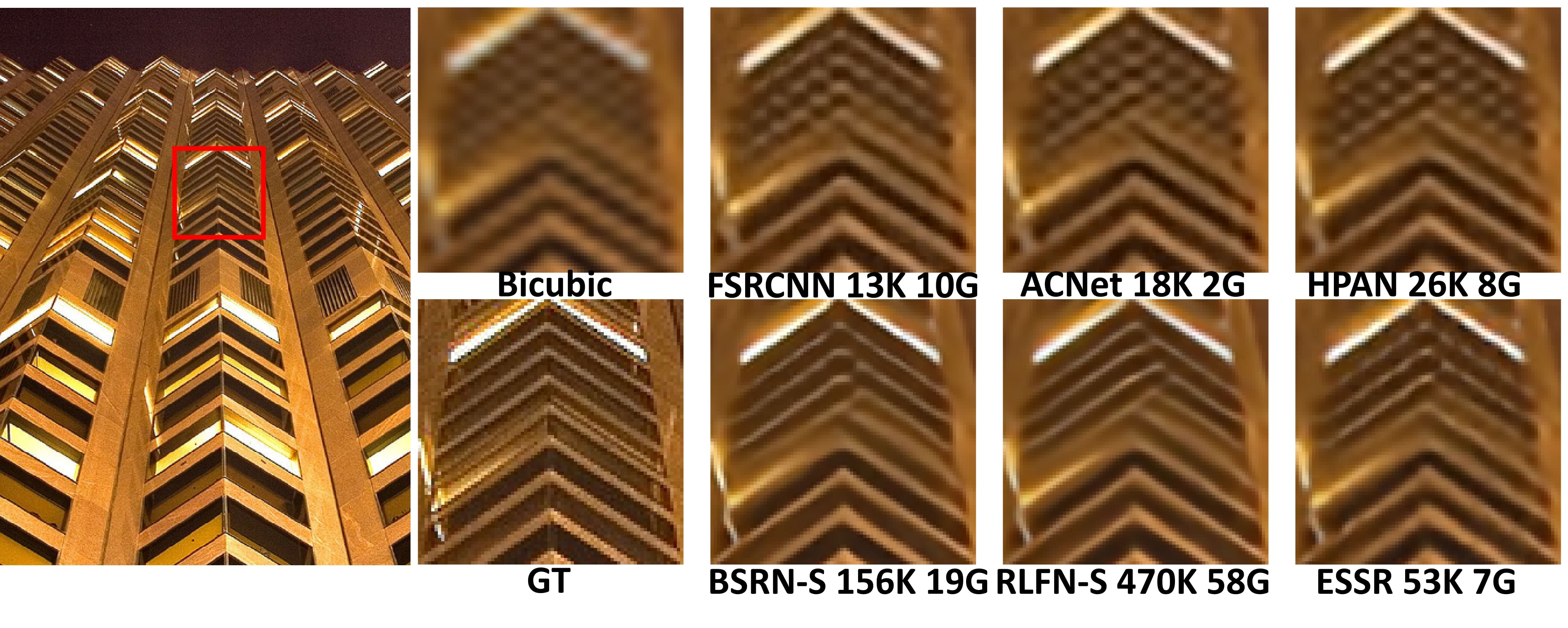}
        \hfill
        \includegraphics[width=0.48\linewidth,keepaspectratio=true]{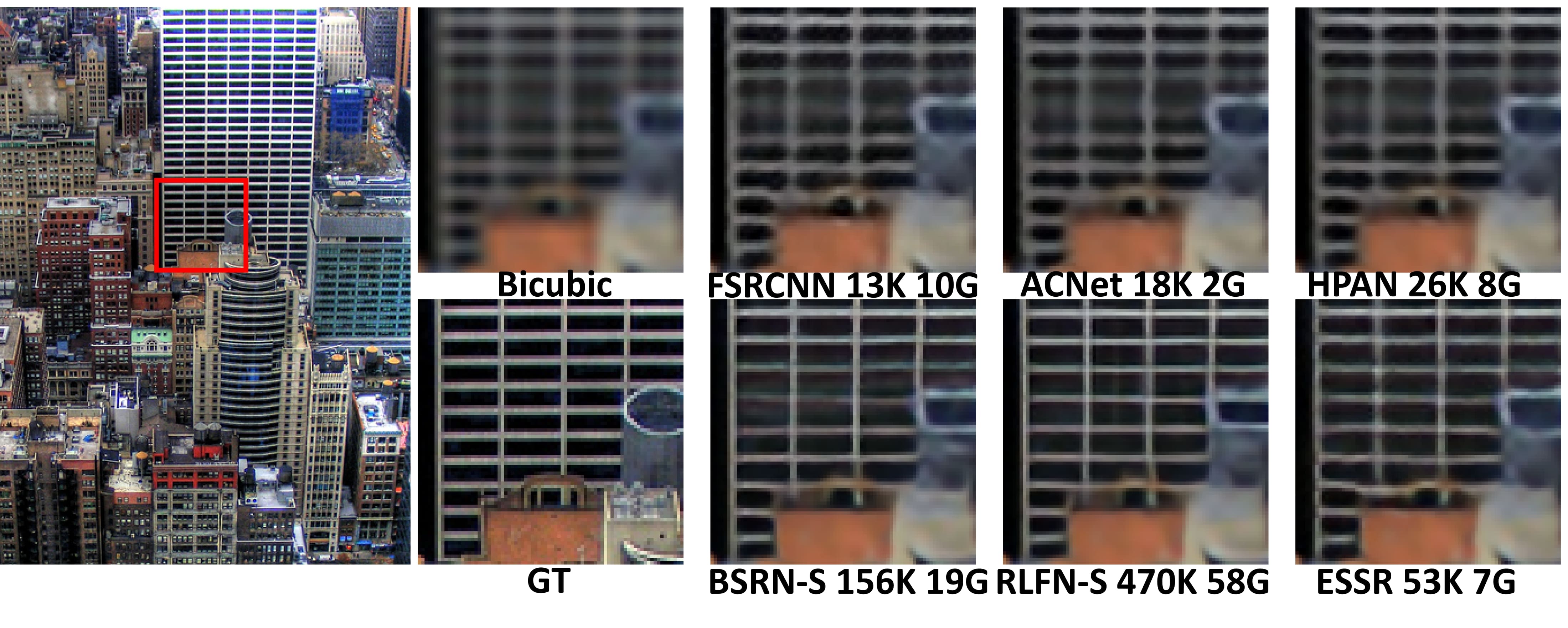}
        \caption{Visual comparison of the PSNR-oriented ESSR with the state-of-the-art methods. (Parameter/MACs). \textbf{(left)} $\times$4 Urban100 img015. \textbf{(right)} $\times$4 Urban100 img073.}
        \label{SW_Image_Comparison_X4_456}
    \end{minipage}

    \begin{minipage}{\textwidth}
        \centering
        \includegraphics[width=0.48\linewidth,keepaspectratio=true]{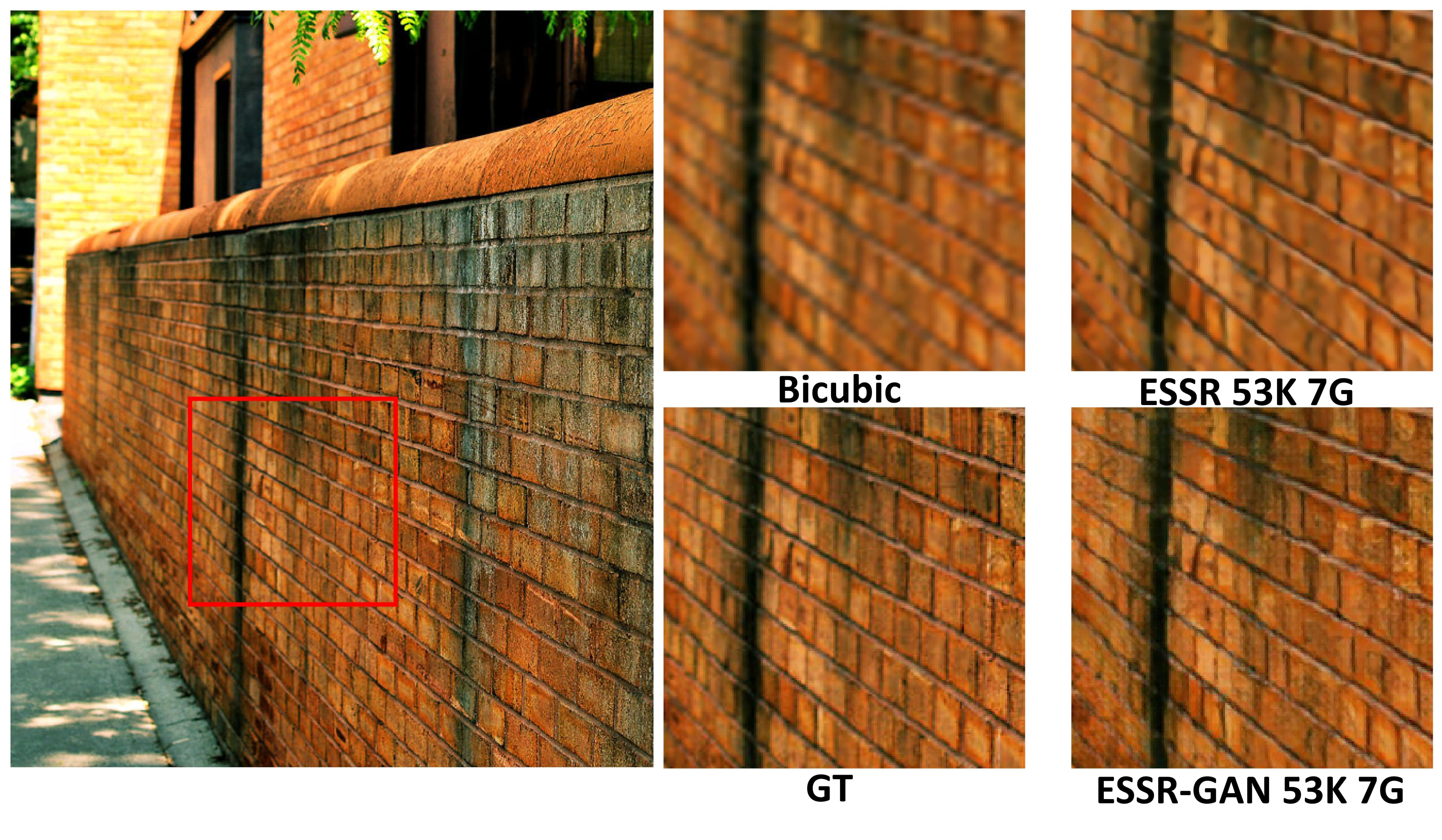}
        \hfill
        \includegraphics[width=0.48\linewidth,keepaspectratio=true]{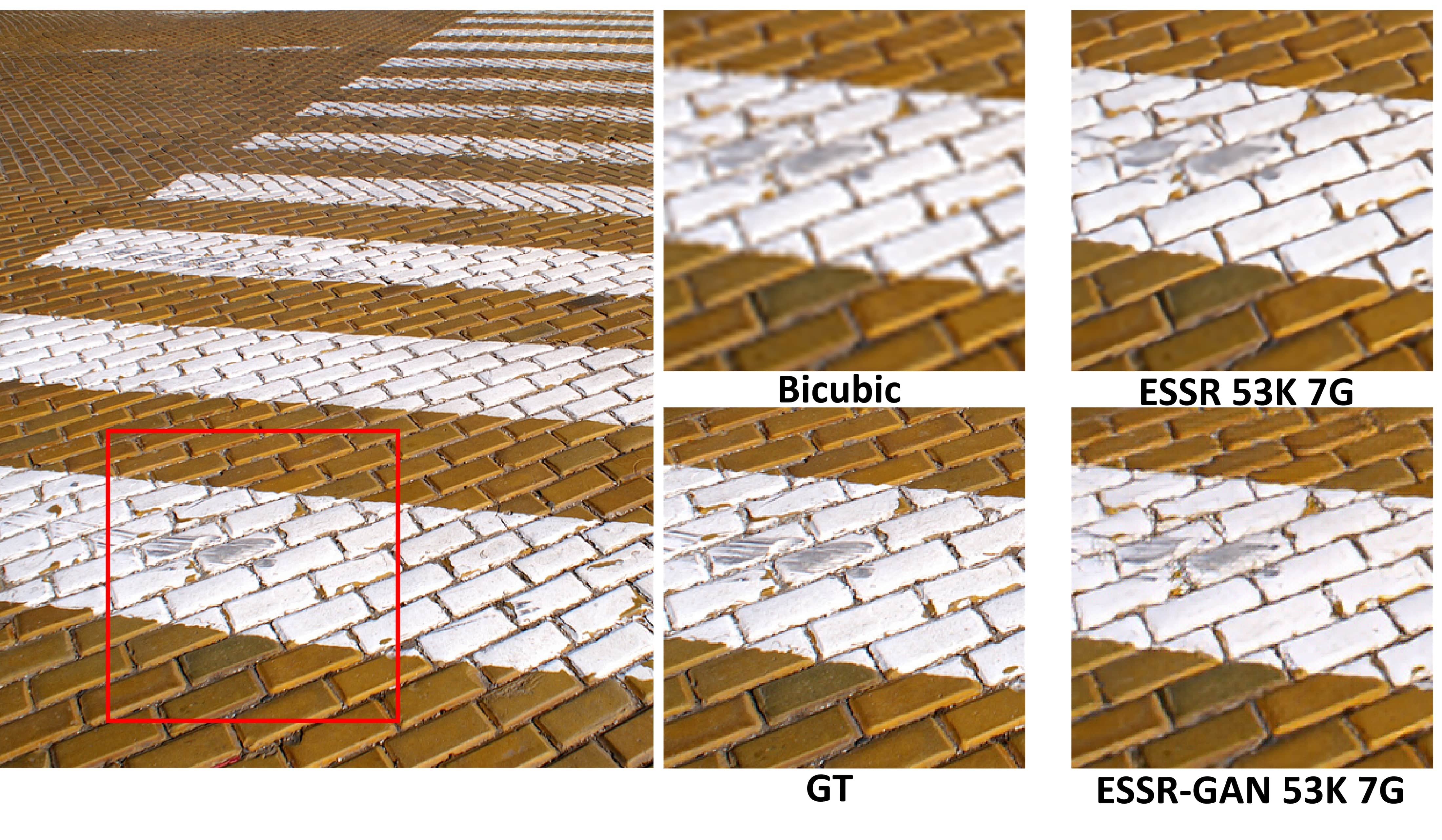}
        \caption{Visual comparison of PSNR-oriented ESSR with perceptual-oriented ESSR. (Parameter/MACs). \textbf{(left)} $\times$4 Urban100 img018. \textbf{(right)} $\times$4 Urban100 img091.}
        \label{GAN_Image_Comparison_X4_12}
    \end{minipage}
    
\end{figure*}

The reconstruction results of our proposed ESSR model are depicted in Fig.~\ref{SW_Image_Comparison_X2_123} to Fig.~\ref{SW_Image_Comparison_X4_123}. At a ×2 scale, ESSR offers clearer image quality and sharper edges than FSRCNN, ACNet, and HPAN. Compared to larger models like BSRN and RLFN-S, our model provides almost the same image quality, even though BSRN and RLFN-S employ more complex methods and larger model sizes. Similar results can be found at a ×4 scale. The advantage of using ESSR becomes even more pronounced. ESSR is capable of reconstructing straight lines, a task that FSRCNN, ACNet, and HPAN struggle with. This inability results in a significant degree of blur and aliasing in their results. This clear distinction highlights the effectiveness of the ESSR model in SR tasks, providing high-quality image reconstructions even at higher scales.

\subsubsection{GAN-based ESSR results}
The results of our GAN-based method are presented in Table~\ref{tab: ×4 scale GAN quantitative results}. Due to our ESSR-GAN being the lightest model, there exists a gap in terms of LPIPS performance between ESSR-GAN and the state-of-the-art methods with parameters in the millions. However, as indicated by the PSNR and SSIM, we are still able to maintain a significant structural similarity with the ground truth.

Fig.~\ref{GAN_Image_Comparison_X4_12} offers a visual comparison between PSNR-oriented ESSR and perceptual-oriented ESSR. The images produced by ESSR-GAN are crisper than those from ESSR, with the edges appearing significantly sharper and exhibiting more textures.

\begin{table*}
\centering
\caption{Quantitative comparison of different methods for the perceptual-oriented ×4 upsampling. ↑: higher is better, ↓: lower is better.}
\label{tab: ×4 scale GAN quantitative results}
\begin{tblr}{
  width = \linewidth,
  colspec = {Q[116]Q[47]Q[62]Q[58]Q[63]Q[62]Q[58]Q[63]Q[62]Q[58]Q[63]Q[62]Q[58]Q[63]},
  cells = {c},
  cell{1}{1} = {r=2}{},
  cell{1}{2} = {r=2}{},
  cell{1}{3} = {c=3}{0.183\linewidth},
  cell{1}{6} = {c=3}{0.183\linewidth},
  cell{1}{9} = {c=3}{0.183\linewidth},
  cell{1}{12} = {c=3}{0.183\linewidth},
  vlines,
  hline{1,3-10} = {-}{},
  hline{2} = {3-14}{},
}
\textbf{Model} & \textbf{Size} & \textbf{Set5} &  &  & \textbf{Set14} &  &  & \textbf{B100} &  &  & \textbf{Urban100} &  & \\
 &  & \textbf{PSNR↑} & \textbf{SSIM↑} & \textbf{LPIPS↓} & \textbf{PSNR↑} & \textbf{SSIM↑} & \textbf{LPIPS↓} & \textbf{PSNR↑} & \textbf{SSIM↑} & \textbf{LPIPS↓} & \textbf{PSNR↑} & \textbf{SSIM↑} & \textbf{LPIPS↓}\\
SRGAN & 1.5 & 29.182 & 0.842 & 0.094 & 26.171 & 0.701 & 0.172 & 25.447 & 0.648 & 0.206 & 24.393 & 0.728 & 0.158\\
ESRGAN & 16.7 & 30.459 & 0.852 & 0.083 & 26.283 & 0.698 & 0.139 & 25.288 & 0.649 & 0.168 & 24.350 & 0.733 & 0.125\\
Real-ESRGAN & 16.7 & 26.617 & 0.807 & 0.169 & 25.421 & 0.696 & 0.234 & 25.089 & 0.653 & 0.282 & 22.671 & 0.686 & 0.214\\
AGD & 0.42 & 30.432 & 0.861 & 0.097 & 27.276 & 0.739 & 0.160 & 26.219 & 0.688 & 0.214 & 24.732 & 0.743 & 0.170\\
EdgeSRGAN & 0.66 & 29.487 & 0.837 & 0.095 & 26.814 & 0.715 & 0.176 & 25.543 & 0.644 & 0.210 & 24.268 & 0.716 & 0.170\\
EdgeSRGAN-tiny & 0.09 & 28.074 & 0.803 & 0.146 & 26.001 & 0.702 & 0.242 & 25.526 & 0.658 & 0.292 & 23.332 & 0.672 & 0.269\\
ESSR-GAN & 0.05 & 30.512 & 0.861 & 0.181 & 27.454 & 0.745 & 0.263 & 26.552 & 0.696 & 0.289 & 24.642 & 0.735 & 0.276
\end{tblr}
\end{table*}

\subsubsection{Comparison with other dynamic methods}
Table~\ref{tab: Dynamic method comparison} offers a comparison of various dynamic methods and their respective costs. In terms of the decision method, unlike SRNPU \cite{SRNPU}, CDNSR~\cite{wang2023classification} and ClassSR \cite{classsr}, which use an additional neural network for prediction, our input edge threshold method is the simplest. For each subnet, both the ARM \cite{ARM} and our method share the same model weight. Overall, we have the lowest hardware cost and are the only ones implementing the whole model on hardware. SRNPU only implements the subnet without the decision network. Meanwhile, CDNSR, ClassSR and ARM propose only software algorithms.

\begin{table}
\centering
\caption{Comparison of the dynamic methods.}
\label{tab: Dynamic method comparison}
\begin{tblr}{
  width = \linewidth,
  colspec = {Q[229]Q[231]Q[167]Q[133]Q[163]},
  cells = {c},
  hlines,
  vline{2-5} = {-}{},
  hline{1,6} = {-}{0.08em},
}
\textbf{Model}  & \textbf{SRNPU \cite{SRNPU}} & \textbf{ClassSR \cite{classsr}} & \textbf{ARM \cite{ARM}} & \textbf{ESSR} \\ \hline
{Decision \\method} & {Decision\\Net}          & {Class\\Module}   & {Lookup \\table} & {Edge \\threshold} \\
Subnet              & Individual               & Individual        & Shared           & Shared             \\
Decision cost       & High                     & High              & Middle           & Low                \\
{HW\\design}        & {Yes(w/o\\~DecisionNet)} & No                & No               & Yes                
\end{tblr}
\end{table}

We present the experimental results comparing our ESSR approach with SRNPU in Table~\ref{ESSR vs SRNPU}. In our case, we showcase two results for ESSR, each employing a different combination of thresholds. The upper result demonstrates nearly identical MACs saving with a lower drop in PSNR. On the other hand, the lower result exhibits almost the same PSNR drop as SRNPU but achieves significantly higher MACs saving.
It is worth noting that in the Manga109 dataset, SRNPU achieves only 5.3\% MACs saving but experiences a PSNR drop of 0.93dB. On the contrary, our ESSR approach can achieve 14\% MACs saving with a mere 0.03dB PSNR drop, or a PSNR drop of 0.94dB with 50.4\% MACs saving. This is due to our effective decision method.

\begin{table}[tb]
\centering
\caption{The comparison of ESSR vs SRNPU on the ×2 scale. ESSR X+Y is ESSR with $threshold_{1} = X$ and $threshold_{2} = Y$}
\label{ESSR vs SRNPU}
\begin{tabular}{c|c|c|c}
\hline
\textbf{Dataset}                  & \textbf{Model} & \textbf{MACs Saving} & \textbf{PSNR Drop} \\ \hline
\multirow{3}{*}{B100}             & SRNPU          & 36.0\%               & -0.26                   \\
                                  & ESSR 15+60     & 34.2\%               & -0.20                   \\
                                  & ESSR 8+80      & 46.7\%               & -0.29                   \\ \hline
\multirow{3}{*}{Urban100}         & SRNPU          & 25.6\%               & -0.62                   \\
                                  & ESSR 8+60      & 27.6\%               & -0.27                   \\
                                  & ESSR 15+80     & 39.5\%               & -0.57                   \\ \hline
\multirow{3}{*}{Manga109}         & SRNPU          & 5.3\%                & -0.93                   \\
                                  & ESSR 8+20      & 14.0\%               & -0.03                   \\
                                  & ESSR 8+80      & 50.4\%               & -0.94                   \\ \hline
\multirow{3}{*}{DIV2K Validation} & SRNPU          & 34.1\%               & -0.38                   \\
                                  & ESSR 8+40      & 36.2\%               & -0.11                   \\
                                  & ESSR 8+80      & 56.8\%               & -0.42                   \\ \hline
\end{tabular}
\end{table}

Table~\ref{table: ESSR vs ARM} shows the comparison with ARM~\cite{ARM} and CDNSR~\cite{wang2023classification} with different backbones. We can achieve better quality across different MACs savings, even for ARM and CDNSR with much larger models.

\begin{table}[tb]
\centering
\caption{The ×4 scale comparison between ARM, CDNSR, and ESSR on Test8K for different MAC reduction rates.}
\label{table: ESSR vs ARM}
\begin{tabular}{@{}llccc@{}}
\toprule
\multicolumn{2}{c}{Model} & \multicolumn{3}{c}{Test8K PSNR(dB)} \\
\cmidrule(r){1-2} \cmidrule(l){3-5}
Model       & Parameters & 40\%  & 50\%  & 60\%  \\ 
\midrule
ARM FSRCNN  & 25K        & 32.75            & 32.73            & 32.66            \\
ARM CARN    & 295K       & 33.31            & 33.27            & 33.18            \\
ARM SRResNet& 1.5M       & 33.52            & 33.50            & 33.46            \\
CDNSR FSRCNN& 25K        & 32.74            & 32.74            & 32.69            \\
CDNSR CARN  & 295K       & 33.15            & 33.15            & 33.15            \\
CDNSR SRResNet& 1.5M     & 33.51            & 33.50            & 33.49            \\
CDNSR RCAN  & 15.6M      & 33.74            & 33.73            & 33.72            \\
Threshold ESSR& 53K      & 34.61            & 34.56            & 34.50            \\
\bottomrule
\end{tabular}
\end{table}

\subsection{Hardware Implementation and Comparison}
The proposed design was developed using Verilog and synthesized with the Synopsys Design Compiler using the TSMC 28nm CMOS technology. Power consumption was measured using Synopsys PrimeTime PX. With the proposed approach, this design can achieve 8K@30FPS using an 800MHz working frequency with a gate count of 2749K, which includes edge threshold computing as well. Our approach also keeps the average power consumption low at just 0.2075W, while energy efficiency peaks at 4797Mpixels/J.
Table~\ref{tab: HW comparison} shows the implementation result and comparisons. 
Overall, our GLNPU shows the highest PSNR on Set5 ×2 scale and the highest throughput among the compared works. 

Compared to non-dynamic processing approaches,  eCNN~\cite{ECNN}, with its computationally demanding model and substantial SRAM requirement for weight storage, lags behind our lightweight and dynamic processing strategy. Specifically, we offer a 24-fold smaller PE count, a seven-fold smaller SRAM, and a marginally superior PSNR. Compared to ACNPU, although GLNPU has nearly three times the number of PEs, GLNPU has a similar gate count due to FXP10 instead of floating points in ACNPU~\cite{ACNPU}. Furthermore, our GLNPU achieves four times throughput and improves performance by 0.3dB. 

Compared to dynamic processing methods, SRNPU \cite{SRNPU} and SRSoC \cite{SRNPU2022} use the much simpler model, FSRCNN, as the backbone.   While this translates to reduced complexity and power consumption, it also results in a diminished PSNR. Their focus on FHD output further conserves power. Additionally, hardware utilization of SRNPU is low, particularly for the small model branch (31.2\%). In contrast, our \textit{configurable group of layer mapping} can synergize with the SFB, ensuring high PE utilization even while running smaller models.

\begin{table*}
\centering
\caption{Performance comparison of super-resolution accelerators}
\label{tab: HW comparison}
\begin{tabular}{c|c|c|c|c|c}
\hline
\textbf{}                     & \textbf{GLNPU} & \textbf{ACNPU \cite{ACNPU}} & \textbf{eCNN \cite{ECNN}}  & \textbf{SRNPU \cite{SRNPU}}     & \textbf{SRSoC \cite{SRNPU2022}} \\ \hline
\textbf{Process}              & 28nm   & 40nm          & 40nm           & 65nm                        & 65nm                   \\ \hline
\textbf{Supply Voltage}       & 0.9V   & 0.9V          & 0.9V           & 1.1V                        & 1.0V                   \\ \hline
\textbf{Algorithm}            & ESSR   & Asymmetric SR & ERNet          & Tile-based Selective FSRCNN & FSRCNN/ClassSR         \\ \hline
\textbf{Precision(W/A)(bits)} & FXP10  & FP10 / FP13   & FXP8           & FXP8 / FXP16                & FXP8 / FXP8 + 5\% FP8  \\ \hline
\textbf{×2 Set5 PSNR(dB)}     & \textbf{37.64}  & 37.30         & 37.62          & 37.06                       & 36.80                 \\ \hline
\textbf{Parameter(K)}         & 51     & 17            & 202            & 183                         & -                      \\ \hline
\textbf{Supported Scale}      & ×2, ×4 & ×2, ×4        & ×2, ×4         & ×2, ×3, ×4                  & ×2, ×4                 \\ \hline
\textbf{Frequency(MHz)}       & 800    & 270           & 250            & 200                         & 200                    \\ \hline
\textbf{Gate Count(K)}        & 2749   & 2332          & -              & -                           & -                      \\ \hline
\textbf{\#PE}                 & 3402   & 1248          & 81920          & 1152                        & -                      \\ \hline
\textbf{SRAM(KB)}             & 388    & 198           & 2864           & 572                         & 177                    \\ \hline
\textbf{Frame Rate(FPS)} &
  \begin{tabular}[c]{@{}c@{}}30(×2, 4K)\\      \textbf{30(×4, 8K)}\end{tabular} &
  \begin{tabular}[c]{@{}c@{}}31.7(×2, FHD)\\      124.4(×4, FHD)\end{tabular} &
  \begin{tabular}[c]{@{}c@{}}30/60(×2/×4, HD)\\      30(×2/×4, 4K)\end{tabular} &
  \begin{tabular}[c]{@{}c@{}}31.8(×2, FHD)\\       88.3(×4, FHD)\end{tabular} &
  \begin{tabular}[c]{@{}c@{}}55(×2, FHD)\\      107(×4, FHD)\end{tabular} \\ \hline
\textbf{Power(W)}             & 0.2075 & 0.1318        & 7.23$\sim$7.46 & 0.211                       & 0.098                  \\ \hline
\textbf{\begin{tabular}[c]{@{}c@{}}SR   Throughput\\      (Mpixels/s)\end{tabular}} &
  \begin{tabular}[c]{@{}c@{}}\textbf{248.8(×2)}\\      \textbf{995.3(×4)}\end{tabular} &
  \begin{tabular}[c]{@{}c@{}}65.7(×2)\\       258.0(×4)\end{tabular} &
  \begin{tabular}[c]{@{}c@{}}248.8(×2)\\       248.8(×4)\end{tabular} &
  \begin{tabular}[c]{@{}c@{}}65.9(×2)\\       183.1(×4)\end{tabular} &
  \begin{tabular}[c]{@{}c@{}}114.0(×2)\\       221.9(×4)\end{tabular} \\ \hline
\textbf{\begin{tabular}[c]{@{}c@{}}Power   Efficiency\\      (Mpixels/s/W)\end{tabular}} &
  \begin{tabular}[c]{@{}c@{}}\textbf{1199.0(×2)}\\      \textbf{4796.6(×4)}\end{tabular} &
  \begin{tabular}[c]{@{}c@{}}498.5(×2)\\      1957.5(×4)\end{tabular} &
  \begin{tabular}[c]{@{}c@{}}33.4(×2)\\      33.4(×4)\end{tabular} &
  \begin{tabular}[c]{@{}c@{}}312.3(×2)\\       867.8(×4)\end{tabular} &
  \begin{tabular}[c]{@{}c@{}}1163.3(×2)\\      2264.3(×4)\end{tabular} \\ \hline
\textbf{\begin{tabular}[c]{@{}c@{}}$^*$Power   Efficiency\\      (Mpixels/s/W)\end{tabular}} &
  \begin{tabular}[c]{@{}c@{}}1199.0(×2)\\      4796.6(×4)\end{tabular} &
  \begin{tabular}[c]{@{}c@{}}712.1(×2)\\      2796.4(×4)\end{tabular} &
  \begin{tabular}[c]{@{}c@{}}49.2(×2)\\      47.6(×4)\end{tabular} &
  \begin{tabular}[c]{@{}c@{}}1083.1(×2)\\       3009.3(×4)\end{tabular} &
  \begin{tabular}[c]{@{}c@{}}3333.9(×2)\\      6489.4(×4)\end{tabular} \\ \hline
\end{tabular}
\\
$^*$Process normalized to 28nm 0.9V. \\
\end{table*}

\subsection{Hardware analysis}
\subsubsection{Area and power analysis}
\begin{figure}[tb]
    \centering
    \includegraphics[height=!,width=1\linewidth,keepaspectratio=true]{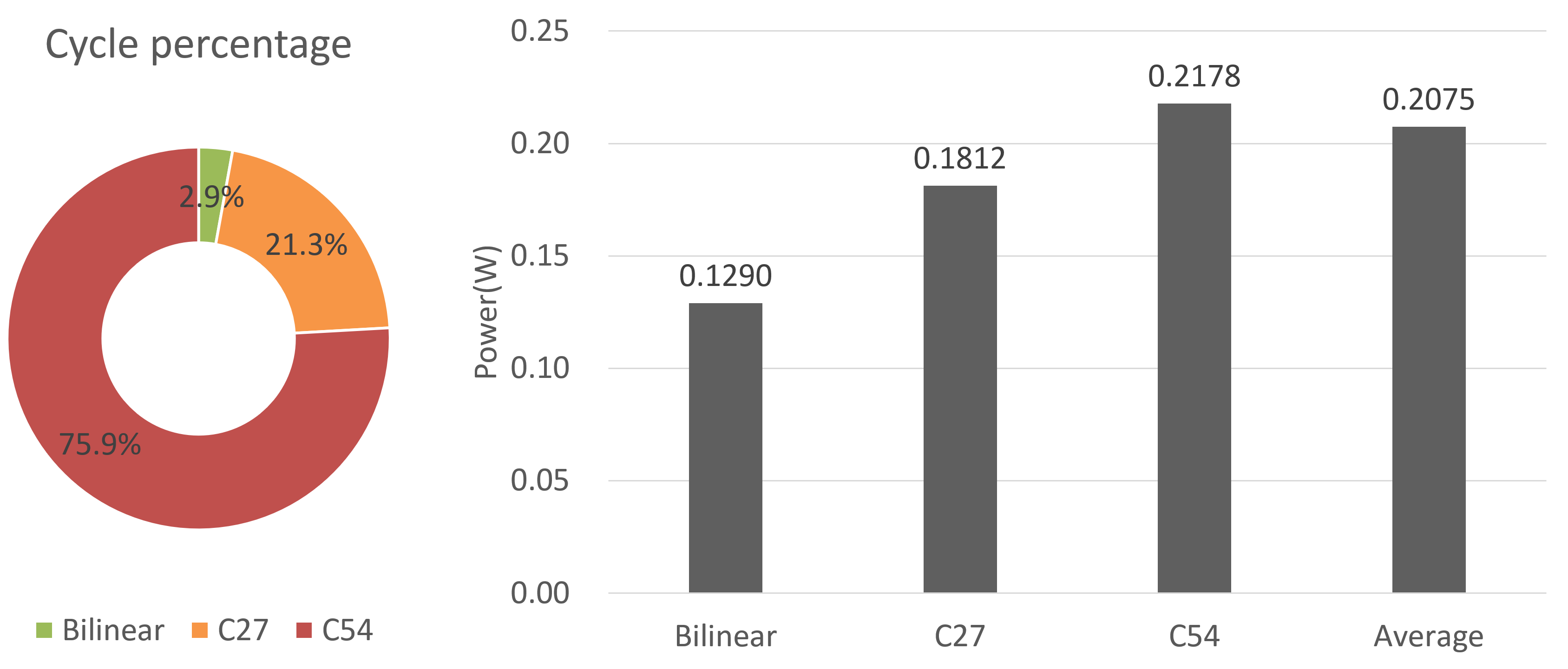}
    \caption {The power analysis of the GLNPU.}
    \label{HW_Power_analysis}
\end{figure}

Fig.~\ref{HW_Power_analysis} illustrates the power analysis. The power usage for the Bilinear, C27, and C54 models is 0.1290W, 0.1812W, and 0.2178W, respectively. The cycle percentages of the respective models are 5.6\%, 20.7\%, and 73.8\% for bilinear, C27, and C54. These values were calculated using $threshold_{1} = 8$ and $threshold_{2} = 40$ on the Test8K dataset on the ×4 scale. Considering the cycle percentage of each model, the average power consumption is 0.2075W.

\begin{figure}[tb]
    \centering
    \includegraphics[trim={1cm 1.5cm 1.5cm 1.5cm},clip, height=!,width=0.5\linewidth,keepaspectratio=true]{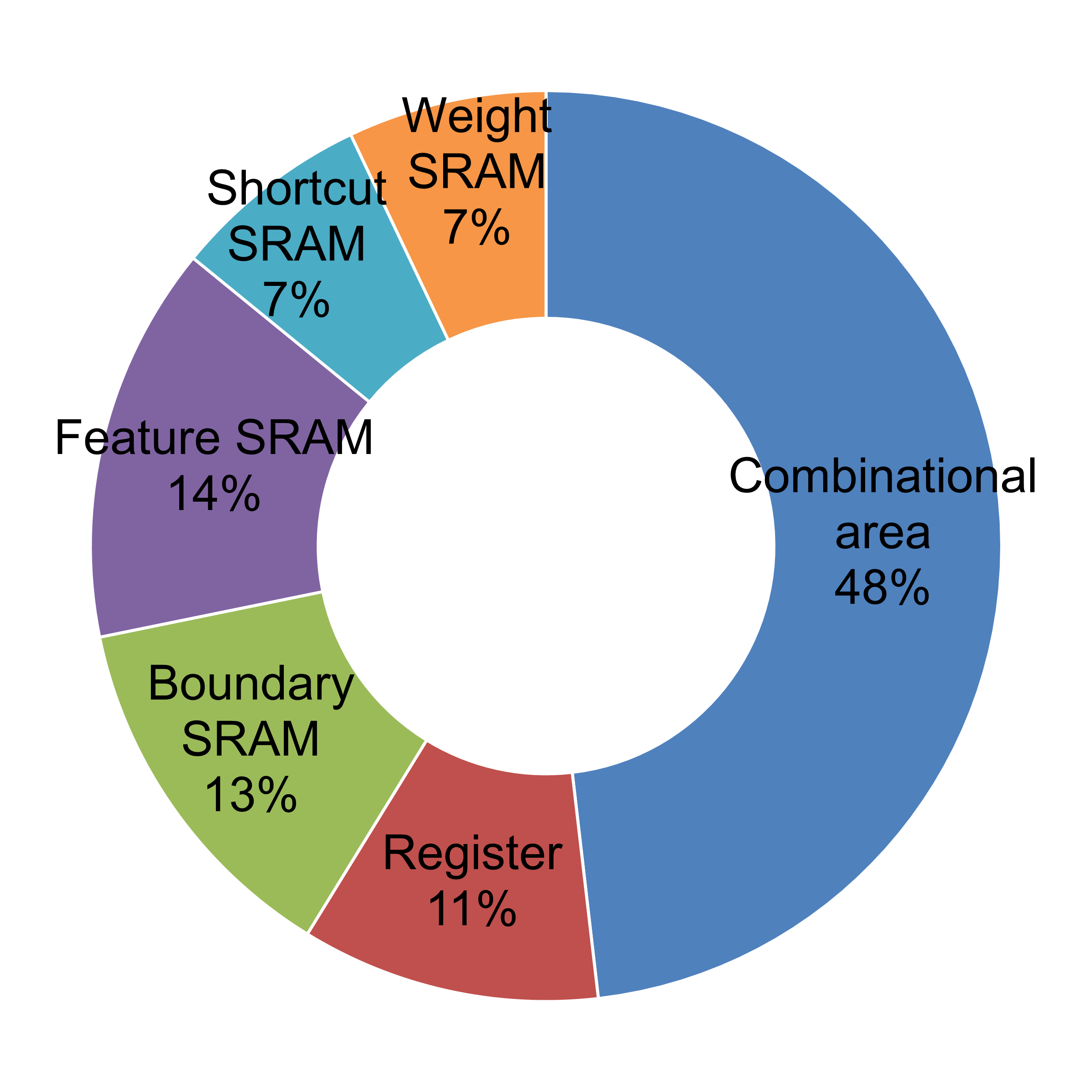}
    \caption {The area breakdown of the GLNPU.}
    \label{HW_area_breakdown}
\end{figure}

Fig.~\ref{HW_area_breakdown} presents the area analysis of our GLNPU. With FXP10 instead of floating points for weight and activation, the PE array and control components account for approximately 59\% of the total area. The SRAM buffers occupy 41\%. The boundary buffer only occupies 13\% due to \textit{Slim overlap block convolution and thick overlap boundary processing} method.

\subsubsection{Boundary processing}
\begin{figure}[tb]
    \centering
    
    \centering
    \includegraphics[height=!,width=1.0\linewidth,keepaspectratio=true]{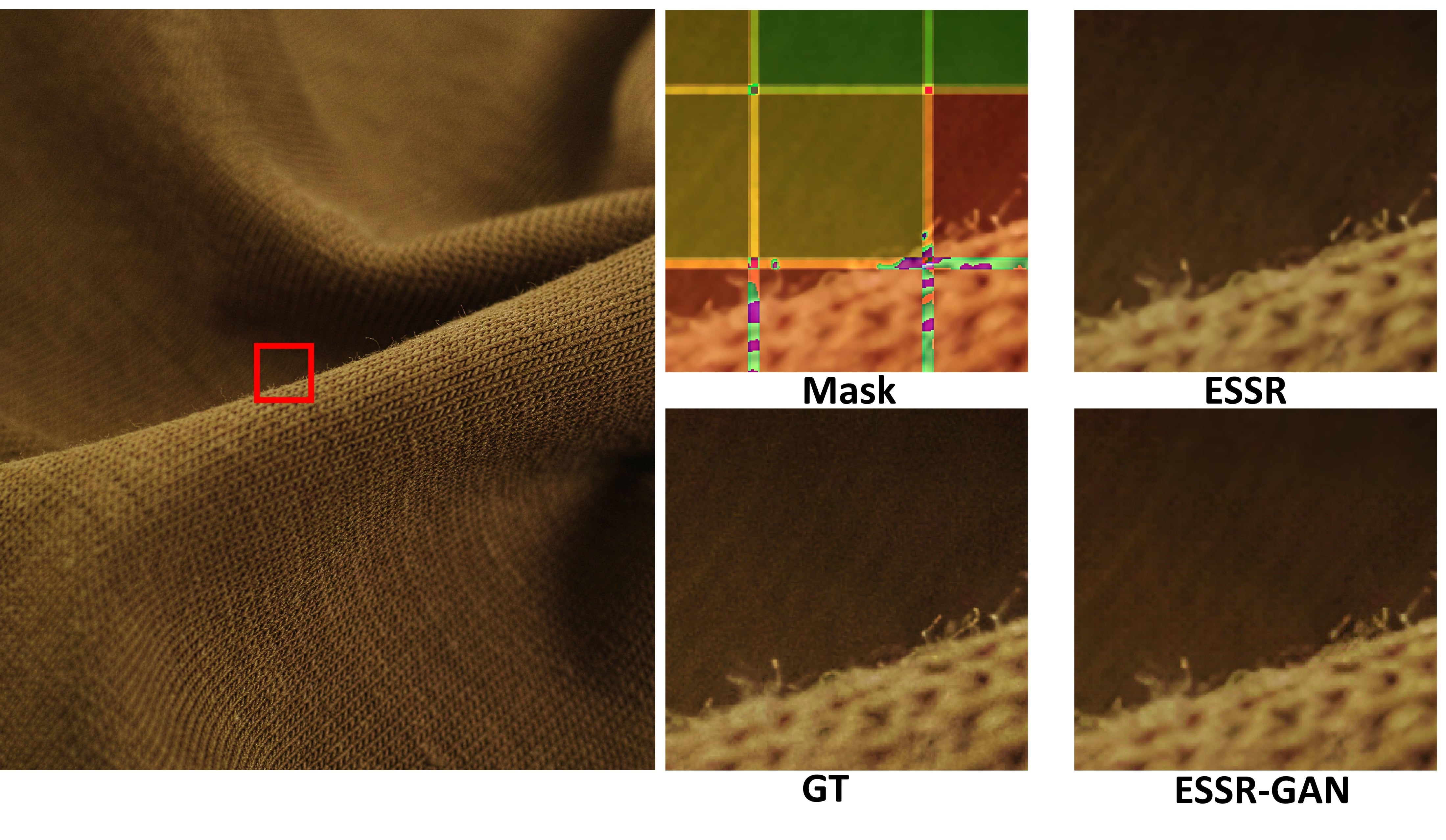}

\caption {Visual results of our boundary solution on the PSNR-oriented ESSR and the perceptual-oriented ESSR for the $\times$4 Test8K 1410.}
\label{Boundary_Image_Comparison_X4_12}
\end{figure}

Fig.~\ref{Boundary_Image_Comparison_X4_12} presents images generated by ESSR and GAN-based ESSR using the \textit{slim overlap block convolution and thick overlap boundary processing} method. The mask indicates which patch employs which subnet: green represents bilinear, yellow is for C27, and red signifies C54. After boundary processing, the discontinuity between pixels in each patch at the ×4 scale is negligible.

\begin{table}{tb}
\centering
\caption{PE utilization of each model for the ×4 scale. }
\label{table:PE utilization}
\begin{tabular}{lllllll} 
\toprule
\textbf{Layer} & \textbf{1x1-A} & \textbf{1x1-B} & \textbf{1x1-C} & \textbf{1x1-D} & \textbf{3x3-B} & \textbf{3x3-C}  \\ 
\midrule
\multicolumn{7}{l}{\textbf{C54}} \\ 
\midrule
BSConv(3,54) & 11\% & 0\% & 0\% & 11\% & 96\% & 96\% \\ 
BSConv(54,54) & 95\% & 95\% & 95\% & 95\% & 95\% & 95\% \\ 
BSConv(54,54) & 95\% & 95\% & 95\% & 95\% & 95\% & 95\% \\ 
1x1(54,54) & 95\% & 95\% & 95\% & 95\% & 0\% & 0\% \\ 
DSConv(54,48) & 96\% & 96\% & 75\% & 75\% & 96\% & 96\% \\ 
\midrule
\multicolumn{7}{l}{\textbf{C27}} \\ 
\midrule
BSConv(3,27) & 11\% & 0\% & 0\% & 0\% & 96\% & 0\% \\ 
SFB & 93\% & 93\% & 93\% & 0\% & 93\% & 93\% \\ 
DSConv(27,48) & 0\% & 96\% & 75\% & 0\% & 96\% & 0\% \\ 
\midrule
\multicolumn{7}{l}{\textbf{Bilinear}} \\ 
\midrule
Bilinear & 0\% & 0\% & 0\% & 0\% & 97\% & 75\% \\
\bottomrule
\end{tabular}

\end{table}

\subsubsection{PE utilization}
The PE utilization for each model as shown in Table~\ref{table:PE utilization} is 15. 3\%, 64. 4\%, and 86. 2\% for bilinear, C27 and C54, respectively. Taking into account the use of the model on the Test8K dataset on the ×4 scale as in Fig.~\ref{HW_Power_analysis}, the average PE utilization is 77.1\%. For the C54 model, utilization of most PEs is 95\% or greater with some idle time due to data loading. The first (BSConv(3, 54) and BSConv(3, 27)) and the last layer (DSConv) have lower utilization due to smaller channel numbers. Bilinear interpolation has the lowest utilization due to smaller channel numbers and simple operations but occupies only 5.6\% cycles.

\section{Conclusion}

This paper introduces an 8K@30FPS SR accelerator with edge-selective dynamic input processing for resource-constrained edge devices. The proposed dynamic processing employs an edge threshold to select subnets for different input complexity, and \textit{resource adaptive model switching} to ensure a balance between minimum and optimal image quality under constraints. The hardware-optimized model, with 51K parameters, achieves a PSNR of 37.64dB on Set5. This represents a 84\% reduction in parameters and a 83\% decrease in MACs, maintaining quality comparable to the baseline model. To maximize hardware efficiency during dynamic processing, this design adopts a \textit{configurable group of layer mapping}. This, in combination with the \textit{structure-friendly fusion block}, achieves 77\% hardware utilization and up to 79\% reduction in SRAM accesses for features. Implemented with the TSMC 28nm process, the design achieves an 8K@30FPS throughput at 800MHz, with an energy efficiency of 4797M pixels/J. The proposed methodology is extendable to other SR network designs.

\bibliographystyle{IEEEtran}

\bibliography{IEEEabrv, bib/thesis}

\begin{IEEEbiography}[{\includegraphics[width=1in,height=1.25in,clip,keepaspectratio]{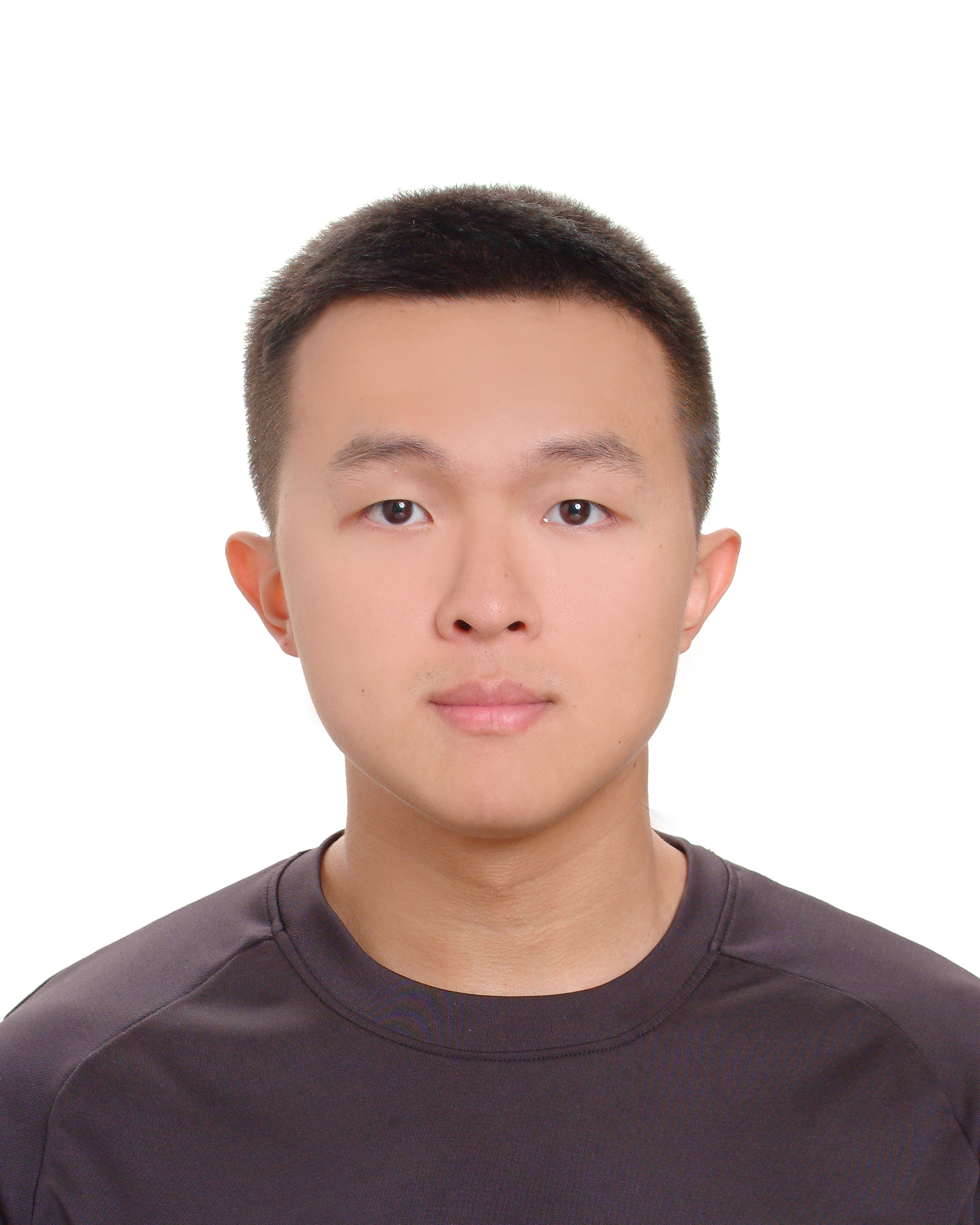}}]{Chih-Chia Hsu}
received the M.S. degree in electronics engineering from the National Yang Ming Chiao Tung University, Hsinchu, Taiwan, in 2023. He is currently working in the MediaTek, Hsinchu, Taiwan. His research interest includes super-resolution neural network and VLSI design.

\end{IEEEbiography}

\begin{IEEEbiography}[{\includegraphics[width=1in,height=1.25in,clip,keepaspectratio]{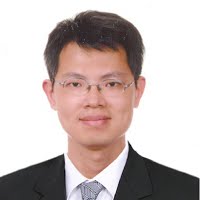}}]{Tian-Sheuan Chang}
	(S’93–M’06–SM’07)
	received the B.S., M.S., and Ph.D. degrees in electronic engineering from National Chiao-Tung University (NCTU), Hsinchu, Taiwan, in 1993, 1995, and 1999, respectively. 
	
	From 2000 to 2004, he was a Deputy Manager with Global Unichip Corporation, Hsinchu, Taiwan. In 2004, he joined the Department of Electronics Engineering, NCTU (as National Yang Ming Chiao Tung University (NYCU) in 2021), where he is currently a Professor. In 2009, he was a visiting scholar in IMEC, Belgium. His current research interests include system-on-a-chip design, VLSI signal processing, and computer architecture.
	
	Dr. Chang has received the Excellent Young Electrical Engineer from Chinese Institute of Electrical Engineering in 2007, and the Outstanding Young Scholar from Taiwan IC Design Society in 2010. He has been actively involved in many international conferences as an organizing committee or technical program committee member.
\end{IEEEbiography}
\end{document}